\documentclass[draft]{agujournal2019}
\usepackage{url} %this package should fix any errors with URLs in refs.
\usepackage{soul}
\draftfalse
\usepackage{lmodern}
\usepackage{graphicx} % Required for inserting images
\usepackage{amsmath} 
\usepackage{amsfonts}
\usepackage{array}
\usepackage{color}
\newcommand{\norm}[1]{\left\lVert#1\right\rVert_2}

\journalname{Water Resources Research}

\begin{document}

\title{Parameter Estimation in River Transport Models With Immobile Phase Exchange Using Dimensional Analysis and Reduced-Order Models}
\authors{Manuel M. Reyna\affil{1}, Alexandre M. Tartakovsky\affil{{1},{2}}}

\affiliation{1}{Department of Civil and Environmental Engineering, University of Illinois Urbana-Champaign, Urbana, IL, USA}
\affiliation{2}{Pacific Northwest National Laboratory,
  Richland, WA 99352}

\correspondingauthor{Alexandre M. Tartakovsky}{amt1998@illinois.edu}

\begin{keypoints}
    \item Dimensionless synthetic datasets allow for efficient parameter estimation of river transport processes and can be reused for new data
    \item A new parameter estimation method is compared with existing methods, providing new insights into their performance and limitations
    \item A dataset of river tracer tests is labeled with transport parameters that can be used for modeling and machine learning 
\end{keypoints}

\begin{abstract}
    We propose a framework for parameter estimation in river transport models using breakthrough curve data, which we refer to as \textit{Dimensionless Synthetic Transport Estimation (DSTE)}. We utilize this framework to parameterize the one-dimensional advection-dispersion equation model, incorporating immobile phase exchange through a memory function. We solve the governing equation analytically in the Laplace domain and numerically invert it to generate synthetic breakthrough curves for different memory functions and boundary conditions. A dimensionless formulation enables decoupling the estimation of advection velocity from other parameters, significantly reducing the number of required forward solutions. To improve computational efficiency, we apply a \textit{Karhunen–Loève (KL) expansion} to transform the synthetic dataset into a reduced-order space. Given a measured breakthrough curve, we estimate the advection velocity by minimizing the distance from the measurement to the synthetic data in KL space, and infer the remaining dimensionless parameters by \textit{Projected Barycentric Interpolation (PBI)}. We benchmark our method against several alternatives, including Laplace domain fitting, moment matching, global random optimization, and variations of the DSTE framework using nearest-neighbor interpolation and neural network-based estimation. Applied to 295 breakthrough curves from 54 tracer tests in 25 rivers, DSTE delivers accurate parameter estimates. The resulting labeled dataset allows researchers to link transport parameters with hydraulic conditions, site characteristics, and measured concentrations. 
    The synthetic dataset can be leveraged for the analysis of new breakthrough curves, eliminating the need for additional forward simulations.
\end{abstract}

\section*{Plain Language Summary}

Understanding how nutrients, pollutants, and other substances move through rivers is important for water quality and environmental management. Tracer tests, consisting of releasing a tracer and measuring its concentration at different downstream location, is a common way of studying transport in rivers. In this study, we conduct a comprehensive analysis of tracer tests to determine the transport properties that influence the velocity of tracer movement, the extent of its dispersion in water, and the duration it remains in stagnant areas of the river, and propose a new parameter estimation method. We analyze 54 field tracer tests with several mathematical models to estimate parameters in these models. Our method leverages the scaling properties of the mathematical models to reduce the number of simulations required for parameter estimation. The method presented in this work, along with the estimated transport properties, can be utilized by scientists to gain a deeper understanding of the transport of substances in rivers.

\section{Introduction}

An accurate understanding of solute transport in river systems is essential for managing water quality, predicting contaminant dispersion, and designing environmental remediation strategies \cite{benedini_water_2013, littlewood_estimating_1992}. The most widely used model for describing river transport is the one-dimensional advection-dispersion equation (ADE), as it allows for simple analytical solutions that can be easily parameterized using breakthrough measurements. However, this model cannot account for long tails often present in the measured breakthrough curves \cite{aubeneau_effects_2015, knapp_perspective_2020}.

The prevailing explanation for this behavior is the interaction between a mobile flow zone and an immobile storage zone. Although models incorporating this dual domain concept are well established (known as transport models with immobile phase exchange or advection-dispersion storage models), there is little consensus on what constitutes the immobile zone: some emphasize dead or recirculation zones \cite{valentine_longitudinal_1977, deng_longitudinal_2002, lees_relationship_2000}, others focus on hyporheic exchange \cite{chen_unifying_2024, aghababaei_temporal_2023, tonina_hyporheic_2007}, and some consider both \cite{seo_moment-based_2001}. These components are often treated as a single black-box process \cite{cardenas_hyporheic_2015}, highlighting the need for improved parameter estimation to better understand river transport dynamics \cite{knapp_perspective_2020}.

The Transient Storage Model (TSM) is the most widely used model for representing immobile exchange in river systems \cite{bencala_simulation_1983, knapp_perspective_2020}. A more general formulation, introduced by \citeA{haggerty_late-time_2000}, models exchange with the immobile phase as a convolution with a memory function that characterizes the residence time distribution. Analytical expressions for the temporal moments of this model have been developed for both general and specific memory functions \cite{aghababaei_temporal_2023}, and have been applied to parameter estimation in river transport studies, including those using the TSM \cite{seo_moment-based_2001, gonzalez-pinzon_scaling_2013}. Alternative methods include fitting Laplace-transformed breakthrough curves to analytical solutions in the transformed domain \cite{seo_moment-based_2001}, matching non-moment-based statistics \cite{runkel_new_2002}, or manual curve fitting \cite{nordin_empirical_1974}. These methods often suffer from low precision, sensitivity to model misspecification, and limited applicability.

To overcome the limitations of traditional parameter estimation methods, we introduce a new framework—Dimensionless Synthetic Transport Estimation (DSTE)—that integrates dimensional analysis, reduced-order modeling, and interpolation on the synthetic manifold to estimate transport model parameters from breakthrough curve data. Dimensional analysis, a well-established tool in transport modeling, reduces the number of governing variables by identifying dimensionless groups \cite{bremer_bodenstein_2024}, which in turn reduces the number of input variables of surrogate models \cite{hazyuk_optimal_2017}. DSTE consists of creating a synthetic dataset of breakthrough curves obtained by solving the dimensionless transport equations and then fitting measurements to them to infer transport parameters.

To further reduce computational cost and complexity (making interpolation more tractable), we apply reduced-order modeling. Among dimensionality reduction techniques, linear methods like Karhunen–Loève (KL) decomposition offer attractive properties, including interpretability, computational efficiency, reduced risk of overfitting, and a natural link to least-squares minimization \cite{palo_dimensionality_2021, bhattacharya_model_2021}.

The synthetic dataset consists of a number of solutions of the governing equations, generated using parameters sampled from a distribution derived from estimates obtained by Laplace-domain fitting and moment matching. To compute these solutions, we apply the Laplace transform to the governing dimensionless partial differential equation, converting it into an ordinary differential equation—a well-established technique to solve transport models \cite{toride_comprehensive_1993}. When analytical inverse transforms are unavailable or impractical due to complexity, numerical Laplace inversion provides an efficient and accurate alternative \cite{wang_surface_2021}. In this work, we combine analytical forward Laplace transforms with numerical inverse transforms to solve the advection-dispersion equation with immobile phase exchange.

We compare four approaches for interpolation on a synthetic manifold. Ordered by increasing complexity, these methods include  Nearest Neighbor Interpolation (NNI), a new method that we term Projected Barycentric Interpolation (PBI), and two Deep Neural Network (DNN) models—one trained directly for inverse mapping, and another trained for forward mapping followed by inversion. DNNs have proven effective as surrogates for complex models due to their function approximation capabilities and differentiability, which is particularly useful for inverse problems. In particular, Karhunen-Loève Deep Neural Networks (KL-DNNs)—which operate on KL expansion coefficients—have shown success in both forward \cite{wang_bayesian_2024, bhattacharya_model_2021} and inverse modeling tasks \cite{wang_total_2024}. In our application of interpolation methods, particular care is taken to ensure that model misspecification (also known as model inadequacy or mismatch) is handled correctly, so that the target of our estimation is the projection of the data onto the model \cite{park2025robustuniversalinferencemisspecified}.

In our experiments, PBI and the forward-trained DNN outperformed other methods, with PBI offering a compelling trade-off between accuracy and computational cost. Surprisingly, NNI also performed well, especially considering its simplicity. The primary method presented in this work is therefore DSTE-KL-PBI (the DSTE framework using KL decomposition for reduced-order modeling and Projected Barycentric Interpolation).  For completeness, we also tested ready-made manifold learning and interpolation methods, which underperformed relative to our custom approaches. Finally, we benchmark all methods against a state-of-the-art global optimization technique.

We apply our parameter estimation framework to 295 breakthrough curves and analyze the resulting parameter distributions and fitting errors. This analysis offers insight into the physical characteristics of the systems represented by the data, enabling us to assess the adequacy of the transport models. By comparing the performance of different estimation methods, we also assess their robustness and potential for improving the interpretation of field-scale solute transport. 

The paper is structured as follows. Section \ref{sec:data} introduces the dataset. Section \ref{sec:models} describes the transport models and governing equations. Section \ref{sec:methods} outlines our parameter estimation methodology, and Section \ref{sec:other-methods} reviews comparison methods. Numerical results are presented in Section \ref{sec:results}, and field data analysis follows in Section \ref{sec:data-analysis}. Conclusions are given in Section \ref{sec:conclusions}.

\section{Data}\label{sec:data}

We used a subset of the TIERRAS dataset \cite{rodriguez_tierras_2025}, consisting of 295 field breakthrough curves measured during 54 tracer experiments in 25 locations \cite{nordin_empirical_1974, covino_stream-groundwater_2011, gooseff_comparing_2003, covino_land_2012, rowinski_response_2008, cushing_transport_1993, payn_channel_2009, greenwald_hyporheic_2008}. This part of the dataset encompasses small streams to large rivers, with discharges at the time of measurement ranging from 0.002 to 7,000 cubic meters per second. Cl, Rhodamine B, and Rhodamine WT were used as tracers. Monitoring methods were variable, but included both sample collection and in situ fluorometry. We only analyzed tracer tests in which the release was instantaneous to avoid ambiguities associated with deconvolutions.

\section{Transport Models}\label{sec:models}

We use the one-dimensional advection-dispersion equation with a mass exchange term containing a memory function $g(t)$ to model the considered tracer tests \cite{aghababaei_temporal_2023}:
\begin{linenomath*}\begin{align} \label{eq:memory_ADE_dimensional_system}
    \frac{\partial c}{\partial t} + v \frac{\partial c}{\partial x} - D \frac{\partial^2 c}{\partial x^2} = -\beta  \frac{\partial c_{im}}{\partial t};\quad x\in\Omega,\quad  t \in [0,\infty);
    \\\nonumber
    c_{im} = \int_0^t g(t-\omega) c(x,\omega) d\omega; \quad c = c(x,t),\;c_{im}=c_{im}(x,t),
\end{align}\end{linenomath*}
where $c$ is the concentration in the mobile phase,  $c_{im}$ is the concentration in the immobile phase, $t$ is the time, $x$ is the spatial coordinate, $\beta$ is the ratio between the effective immobile phase area and the mobile phase area, $g(t)$ is a memory function that describes the distribution of residence times in the immobile zone, and $\Omega$ is the spatial domain. 

A similar model was proposed in  \cite{carrera_matrix_1998} for flow in porous media and \cite{haggerty_late-time_2000} for flow in open channels. We consider two common forms of $g(t)$, first-order exchange and power-law, which are listed in Table \ref{tab:memory_functions}.
\begin{table}
    \caption{Memory functions and their Laplace transforms \cite{ginn_phase_2017}. The parameters $k_f$ and $k_r$ are the forward and reverse transfer rates, $\hat{k}_f = k_f \frac{L}{v}$, and $\hat{k}_r = k_f \frac{L}{v}$. $\gamma$ is the main dimensionless parameter of the power-law model. $\alpha$ is a parameter we add for dimensional consistency, with $\hat{\alpha}=\alpha\left(\frac{L}{v}\right)^{1-\gamma}$. The definition of the power-law memory function as a distribution of residence times is not possible because its integral doesn't converge. However, its use in the literature is still ample owing to its connection to exchange into diffusive layers \cite{ginn_phase_2017}.  \textsuperscript{a}Most methods approximate $\hat{g}(0)$ with a truncated power low model.   \cite{haggerty_late-time_2000}.}
    \centering
    \begin{tabular}{>{\raggedright\arraybackslash}p{0.12\linewidth}|>{\raggedright\arraybackslash}p{0.2\linewidth}>{\centering\arraybackslash}p{0.12\linewidth}>{\centering\arraybackslash}p{0.09\linewidth}>{\centering\arraybackslash}p{0.05\linewidth}>{\centering\arraybackslash}p{0.13\linewidth}>{\centering\arraybackslash}p{0.07\linewidth}} 
         Memory function&   E.g. of use&$g(t)$&  $\hat{g}(\hat{t})$&  $\hat{g}(0)$&  $\frac{\partial \hat{g}(\hat{t})}{\partial \hat{t}}$& $\hat{G}(s)$\\ \hline
         First-order&   \citeA{aghababaei_temporal_2023} &$k_f e^{-k_r t}$&  $\hat{k}_f e^{-\hat{k}_r \hat{t}}$&  $\hat{k}_f$&  $-\hat{k}_f\hat{k}_r e^{-\hat{k}_r \hat{t}}$& $\frac{\hat{k}_f}{\hat{k}_r+s}$\\   
         Pure power-law&  \citeA{schumer_fractal_2003}, \citeA{meerschaert_tempered_2008}&$\alpha \frac{t^{-\gamma}}{\Gamma(1-\gamma)}$; $0<\gamma \leq 1$&  $\hat{\alpha} \frac{\hat{t}^{-\gamma}}{\Gamma(1-\gamma)}$&  $\infty$\textsuperscript{a}&  $\hat{\alpha} \frac{\hat{t}^{-\gamma-1}}{\Gamma(-\gamma)}$& $\hat{\alpha}s^{\gamma-1}$\\ 
    \end{tabular}
    \label{tab:memory_functions}
\end{table}

We assume that in all considered experiments, the point release of a tracer occurs instantaneously.  The time and location of the tracer release are denoted as $t=0$ and $x=0$, respectively. The positive $x$ direction coincides with the flow direction. Several initial and boundary conditions can be used to describe such tracer tests. We consider the three most frequently used conditions, listed in Table \ref{tab:problems-comparison}. 

Transport PDE models describing tracer experiments can be defined on semi-infinite $(\Omega=[0,\infty))$ or infinite ($\Omega=(-\infty,\infty)$) domains. In both cases, the tracer release location is set to $x=0$. A PDE model defined on an infinite domain provides a more accurate mathematical description of such tracer tests as it allows for upstream diffusion. 
The choice of the domain type is often made for practical reasons, with the semi-infinite domain being preferred because numerical solutions on it are typically less computationally demanding than those on an infinite domain.

In modeling tracer tests using a semi-infinite domain, two boundary conditions at $x=0$ are commonly used: a Dirichlet boundary condition that neglects upstream dispersion (the boundary condition in formulation 1, Table \ref{tab:problems-comparison}), and a Robin (mixed) boundary condition that accounts for upstream dispersion (formulation 2, Table \ref{tab:problems-comparison}). Both boundary conditions approximate the solution on the infinite domain (formulation 3, Table \ref{tab:problems-comparison}) for $x>0$, and converge to it as the Peclet number increases. In this work, we propose a new boundary condition for the semi-infinite domain (the boundary condition in formulation 4, Table \ref{tab:problems-comparison}) that matches the Laplace-domain solution of the infinite-domain problem for $x>0$. This boundary condition enables the computation of full-domain behavior while retaining the computational advantages of a semi-infinite domain formulation.

\begin{table}
   \caption{Mathematical formulations (definition of the domain and initial and boundary conditions) of a tracer test with the instantaneous point release of a tracer. $\Omega$ is the space domain, $M_0$ is the mass injected by cross-sectional area, and $\hat{M}_0=\frac{M_0}{L}$. Also shown are the function $B(\hat{s})$ in the Laplace space solution of Eq \eqref{eq:laplace-solution} and the solution of ADE without mass exchange (for $g(t)=0$) in the physical space for the considered initial and boundary conditions. 
   \textsuperscript{a}To our knowledge, no analytical inversion exists for these initial and boundary conditions.}
    \centering
    \begin{tabular}{>{\raggedright\arraybackslash}p{0.14\linewidth}|>{\centering\arraybackslash}p{0.17\linewidth}|>{\centering\arraybackslash}p{0.2\linewidth}|>{\centering\arraybackslash}p{0.14\linewidth}|>{\centering\arraybackslash}p{0.19\linewidth}}
         Problem formulation& (1) $\Omega = [0,\infty)$ w/o upst. diff.& (2) $\Omega = [0,\infty)$ w/ upst. diff.&(3) $\Omega = (-\infty,\infty)$& (4) $\Omega = [0,\infty)$ equivalent to (3)\\ \hline
         Initial conditions&   $c(x,0) = 0$&$c(x,0) = 0$ &$c(x,0) = M_0\delta(x)$ &$c(x,0) = 0$\\\hline
         Boundary conditions&   $v c (0,t) = M_0 \delta(t)$&$\left. \left(v c - D \frac{\partial c}{\partial x}\right)\right|_{x=0}$ $= M_0 \delta(t)$&$c(-\infty,t) = 0$ &$\left. \left(v c - 2D \frac{\partial c}{\partial x}\right)\right|_{x=0}$ $= M_0 \delta(t)$\\\hline
 References &  \citeA{gonzalez-pinzon_scaling_2013}, \citeA{aghababaei_temporal_2023}& \citeA{wang_surface_2021}& \citeA{fischer_chapter2_1979}, \citeA{seo_moment-based_2001}&\\\hline
         Initial conditions dimensionless&   $\hat{c}(\hat{x},0) = 0$&$\hat{c}(\hat{x},0) = 0$ &$\hat{c}(\hat{x},0) = \hat{M}_0\delta(\hat{x})$ &$\hat{c}(\hat{x},0) = 0$\\\hline
         Initial conditions dimensionless&   $\hat{c} (0,t) = \hat{M}_0 \delta(\hat{t})$&$ \left.\left( \hat{c} - \frac{1}{Pe} \frac{\partial \hat{c}}{\partial \hat{x}}\right)\right|_{\hat{x}=0}$ $= \hat{M}_0 \delta(\hat{t})$&$\hat{c}(-\infty,\hat{t}) = 0$ &$ \left.\left( \hat{c} - \frac{2}{Pe} \frac{\partial \hat{c}}{\partial \hat{x}}\right)\right|_{\hat{x}=0}$ $= \hat{M}_0 \delta(\hat{t})$\\\hline
 $\hat{c}(0) $& $0$& $0$&$\hat{M}_0$ &$0$\\\hline
 Laplace boundary conditions& $\hat{C}(0,\hat{s}) = \hat{M}_0$;  $\hat{C}(\infty,\hat{s}) = 0$& $ \left.\left( \hat{C}- \frac{1}{Pe} \frac{\partial \hat{c} }{\partial x}\right)\right|_{\hat{x}=0}$ $= \hat{M}_0$; $\hat{C}(\infty,\hat{s}) = 0$&$\hat{C}(-\infty,\hat{s}) = 0$; $\hat{C}(\infty,\hat{s}) = 0$&$ \left.\left( \hat{C}- \frac{2}{Pe} \frac{\partial \hat{c}}{\partial x}\right)\right|_{\hat{x}=0}$ $= \hat{M}_0$; $\hat{C}(\infty,\hat{s}) = 0$\\\hline
 $B(\hat{s})$ \eqref{eq:laplace-solution}& $1$& $\frac{1}{\frac{1}{2}+
\frac{\sqrt{4\left(\hat{s} +  \beta \hat{s}\hat{G}(\hat{s})\right)+Pe}}{2 \sqrt{Pe}}}$&\multicolumn{2}{c}{$\frac{\sqrt{Pe}}{\sqrt{4\left(\hat{s} +  \beta \hat{s}\hat{G}(\hat{s})\right)+Pe}}$}\\\hline
 $\hat{c}(\hat{x},\hat{t})$ for $\hat{g}(\hat{t})=0$ & $\frac{\hat{M}_0\hat{x}}{\sqrt{4\pi\frac{\hat{t}^3}{Pe}}} e^{-\frac{(\hat{x}-\hat{t})^2}{4\frac{\hat{t}}{Pe}}}$& ?\textsuperscript{a}&\multicolumn{2}{c}{$\frac{\hat{M}_0}{\sqrt{4\pi\frac{\hat{t}}{Pe}}}e^{-\frac{(\hat{x}-\hat{t})^2}{4\frac{\hat{t}}{Pe}}}$}\\
    \end{tabular}
    \label{tab:problems-comparison}
\end{table}

We note that Eq \eqref{eq:memory_ADE_dimensional_system} can be rewritten as an integrodifferential equation by applying the Leibniz rule
\begin{linenomath*}\begin{align}\label{eq:memory_ADE_dimensional}
    \frac{\partial c}{\partial t} + v \frac{\partial c}{\partial x} - D \frac{\partial^2 c}{\partial x^2}  &= -\beta\frac{\partial}{\partial t} \int_0^t g(t-\omega) c(x,\omega) d\omega  &
    \\\nonumber
    &= -\beta g(0) c(x,t) -\beta  \int_0^t \frac{\partial g(t-\omega)}{\partial t}  c(x,\omega) d\omega.
\end{align}\end{linenomath*}

Eq \eqref{eq:memory_ADE_dimensional} can be made dimensionless by introducing the characteristic length $L$ (defined as the distance from the tracer source location to point where breakthrough curve measurements are taken) and the dimensionless coordinates  $\hat{x}$, $\hat{t}$, and $\hat{\omega}$  as $L\hat{x}=x$, $\hat{t}\frac{L}{v}=t$, and $\hat{\omega}\frac{L}{v}=\omega$. Then, Eq \eqref{eq:memory_ADE_dimensional} can be rewritten as
\begin{linenomath*}\begin{equation}\label{eq:memory_ADE_dimensionless}
    \frac{\partial \hat{c}}{\partial \hat{t}} + \frac{\partial \hat{c}}{\partial \hat{x}} - \frac{1}{Pe} \frac{\partial^2 \hat{c}}{\partial \hat{x}^2} = -\beta \left(\hat{g}(0) \hat{c}(\hat{x},\hat{t}) + \int_0^{\hat{t}} \frac{\partial \hat{g}(\hat{t}-\hat{\omega})}{\partial \hat{t}} \hat{c}(\hat{x},\hat{\omega}) d\hat{\omega}\right),
\end{equation}\end{linenomath*}
where $c\left(L\hat{x},\hat{t}\frac{L}{v}\right)=\hat{c}\left(\hat{x},\hat{t}\right)$,  $\hat{g}(\hat{\omega}) = \frac{L}{v}g\left(\hat{\omega}\frac{L}{v}\right)$, and $Pe=Lv/D$ is the Peclet number. Eq \eqref{eq:memory_ADE_dimensionless} satisfies $ \hat{c}(\infty,\hat{t}) = 0$. The dimensionless boundary conditions at $x=0$ and the initial conditions are given in Table \ref{tab:problems-comparison}.

\section{Dimensionless Synthetic Transport Estimation (DSTE) Framework With KL Dimension Reduction and Projected Barycentric Interpolation (PBI) of Parameters}\label{sec:methods}

\begin{figure}[htb!]
	\centering
    \includegraphics[angle=0,width=\textwidth]{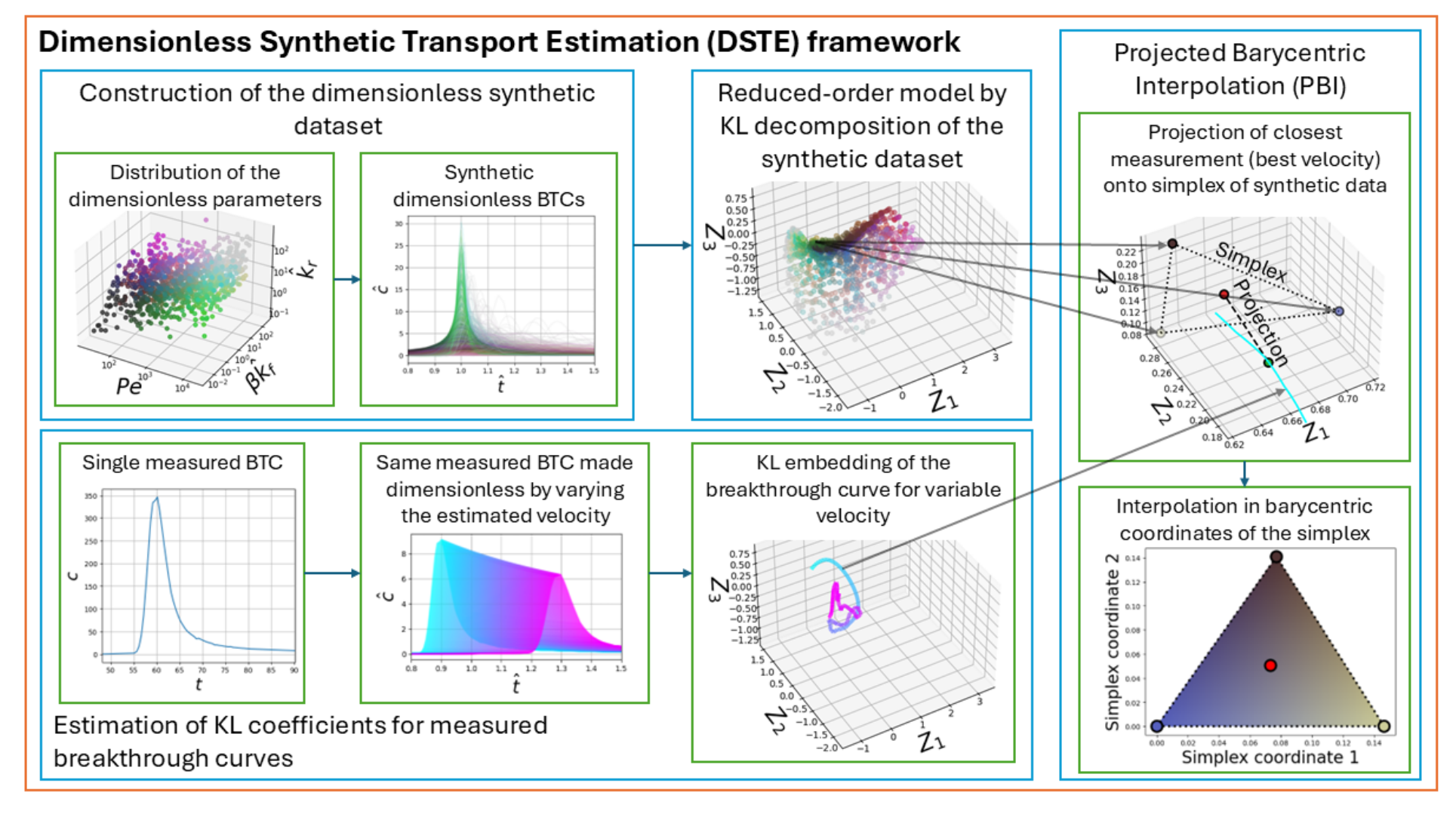}
	\caption{Schematic of the Dimensionless Synthetic Transport Estimation (DSTE) framework with KL reduced-order modeling and Projected Barycentric Interpolation (PBI).}
	\label{fig:method-schematic}
\end{figure}

The DSTE-KL-PBI method for parameter estimation is depicted in Figure \ref{fig:method-schematic}.  The method consists of the following steps:
\begin{enumerate}
    \item \textbf{Construction of the dimensionless synthetic dataset} (Section \ref{sec:methods-synthetic}): A forward solver—using an analytical Laplace-domain solution and a numerical inverse transform (Section \ref{sec:methods-laplace-forward})—is used to generate a set of synthetic dimensionless breakthrough curves given a distribution of dimensionless parameters (which in this work we obtain from ``coarse'' estimates of the parameters by the methods shown in Section \ref{sec:methods-coarse}). This step defines the DSTE framework and offers three significant advantages, (a) it reduces the total number of forward evaluations, (b) it decouples the estimation of the advection velocity from other transport parameters, and (c) the resulting synthetic dataset can be reused across multiple breakthrough curves, allowing us to amortize the computational cost of generating forward solutions.
    \item \textbf{Reduced-order model by KL decomposition of the synthetic dataset} (Section \ref{sec:methods-KL-decomp}): The synthetic breakthrough curves are embedded into a reduced-order space using a Karhunen–Loève (KL) expansion. This reduces computational complexity (it lowers the number of linear interpolations and the size of linear matrix solution needed in step 3) while preserving L2 norms.
    \item \textbf{Estimation of KL coefficients for field breakthrough curves} (Section \ref{sec:methods-Z-MAP}): Measured breakthrough curves are embedded in the KL space using a velocity-dependent transformation. Our approach to estimating the KL coefficients avoids overfitting and tolerates low sampling rates.
    \item \textbf{Projected Barycentric Interpolation (PBI)} (Section \ref{sec:methods-PBI}): The KL-embedded measurements are projected onto the synthetic manifold. The velocity that minimizes the projection distance is selected as the best estimate. Dimensionless parameters are estimated by barycentric interpolation on the simplex formed by the nearest synthetic samples. 
\end{enumerate}

\textit{Note:} The last step can be modified to obtain different methods within the DSTE-KL framework.
 
Although multiple breakthrough curves are available for some tracer experiments, we choose to estimate parameters for the river reaches starting at the point of release and ending at each brintokthrough curve location separately. This is because the parameters vary significantly along the length of the river. Thus, we set $L$ to be the location of a breakthrough curve and compute forward solutions for $c(x=L,t)=\hat{c}(\hat{x}=1,\hat{t})$ rather than the entire concentration field. The estimated parameters represent average transport behavior from the point of tracer release to the point of the concentration measurements. 

The first parameter to be estimated by DSTE methods is the dimensional advection velocity $v$, which we constrain within $0.9 \frac{L}{t_{\text{peak}}^*}<v<1.5\frac{L}{t_{\text{peak}}^*}$, where $t_{\text{peak}}^*$ is the measured time of peak concentration. This range is used for all methods and was selected to ensure that fewer than 5\% of the inferred velocities fall near the imposed bounds. Thremaininghe parameters are made dimensionless and grouped in the vector $\mathbf{y}$ of size $d$. We define the parameter vector $\mathbf{y}=(Pe,\beta \hat{k}_f,\hat{k}_r)^T$ for the first-order memory function model and $\mathbf{y}=(Pe,\beta \hat{\alpha},1-\gamma)^T$ for the power-law memory function model. These sets of parameters are sufficient to represent all possible solutions of Eq \eqref{eq:memory_ADE_dimensional} for the considered memory functions. Note that selecting $\beta$ and either $\hat{k}_f$ or $\hat{\alpha}$ independently only alters the immobile concentration $c_{im}$, which we cannot constrain due to the lack of measurements in the immobile domain.

\subsection{Solver for the Transport Equation for Generating the Training Dataset}\label{sec:methods-laplace-forward}

We generate a synthetic dataset in Step 1 by solving Eq \eqref{eq:memory_ADE_dimensionless} for different values of $\mathbf{y}$ sampled from a random distribution. The considered tracer experiments have $Pe$ between 10 and 40,000. Solving the transport equations using grid-based numerical methods for $Pe>100$ is computationally expensive due to the need for fine spatial and temporal discretization to avoid numerical dispersion and oscillations and to ensure stability \cite{Ferziger2002, zienkiewicz_chapter_2014}.

Transport equations can be solved efficiently for any Peclet number by finding an analytical solution in the Laplace domain and then inverting it analytically \cite{davis_laplace_1985} or numerically \cite{wang_surface_2021}. For the considered Peclet numbers, numerical inversion of the analytical solution in Laplace domain outperforms the finite difference method \cite{runkel_efficient_1993} (even when LU decomposition is used as in \citeA{runkel_OTIS_1998}), and semi-analytical solutions \cite{lassey_unidimensional_1988, lassey_correction_1989, toride_comprehensive_1993}, which require multiple numerical integrations of Bessel functions. Furthermore, solutions for the power-law memory function are available in the Laplace domain but not in the concentration space, since $\hat{g}(0)=\infty$. We numerically invert the Laplace domain solution using the method of de Hoog \cite{de_hoog_improved_1982,wang_surface_2021}. Our algorithm implementation is a Python translation of \cite{invlap}.

The Laplace domain solution is obtained by first taking the Laplace transform of Eq \eqref{eq:memory_ADE_dimensionless}, yielding
\begin{linenomath*}\begin{equation}\label{eq:laplace}
    \hat{s}\hat{C} - \hat{c}(0) + \frac{\partial \hat{C}}{\partial \hat{x}} - \frac{1}{Pe} \frac{\partial^2 \hat{C}}{\partial \hat{x}^2} = -\beta \left(\hat{g}(0) \hat{C} +\left(\hat{s}\hat{G}(\hat{s})-\hat{g}(0)\right)\hat{C}\right),
\end{equation}\end{linenomath*}
where $\hat{C}=\mathcal{L}_{\hat{t}}[\hat{c}](\hat{s})$, $\hat{G}=\mathcal{L}_{\hat{t}}[\hat{g}](\hat{s})$, and $\hat{s}$ is the dimensionless variable in the frequency domain. Rearranging terms in Eq \eqref{eq:laplace}, we obtain the second-order ODE
\begin{linenomath*}\begin{equation}\label{eq:laplace_space_ODE-infinite}
     \left(\hat{s} +  \beta \hat{s}\hat{G}(\hat{s})\right)\hat{C}  + \frac{\partial \hat{C}}{\partial \hat{x}} - \frac{1}{Pe} \frac{\partial^2 \hat{C}}{\partial \hat{x}^2} = \hat{c}(0).
\end{equation}\end{linenomath*}
The domain where the equation is defined, the boundary conditions, and $\hat{c}(0)$ are listed in Table \ref{tab:problems-comparison}. The analytical solution of Eq \eqref{eq:laplace_space_ODE-infinite} defined on $\hat{x}\ge0$ is
\begin{linenomath*}\begin{equation}\label{eq:laplace-solution}
    \hat{C}(\hat{x},\hat{s})= \hat{M}_0 B(\hat{s})  \exp{\left(\frac{\hat{x}}{2}\left(Pe-\sqrt{ Pe\left(4\left(\hat{s} +  \beta \hat{s}\hat{G}(\hat{s})\right)+Pe\right)}\right)\right)},\quad \hat{x}\ge 0;
\end{equation}\end{linenomath*}
 and on the infinite domain $\hat{x}\in(-\infty,\infty)$ it is
\begin{linenomath*}\begin{equation*}
    \hat{C}(\hat{x},\hat{s})=\hat{M}_0 B(\hat{s}) \exp{\left(\frac{Pe\;\hat{x}}{2}-\sqrt{\frac{Pe\;\hat{x}^2}{4}\left(4\left(\hat{s} +  \beta \hat{s}\hat{G}(\hat{s})\right)+Pe\right)}\right)},
\end{equation*}\end{linenomath*}
with $B(\hat{s})$ given in Table \ref{tab:problems-comparison}. 

\subsection{Dimensionless Synthetic Dataset} \label{sec:methods-synthetic}

Because numerical forward solvers are computationally expensive and typically non-differentiable (i.e., the derivatives of the solution with respect to unknown parameters cannot be computed analytically or by means of automatic differentiation), most of the cost in parameter estimation arises from repeated forward simulations. To mitigate this, we precompute a synthetic dataset of dimensionless breakthrough curve solutions.

The synthetic dataset is generated by sampling $\mathbf{y}$ from its prior distribution and solving the governing equation for each sample of $\mathbf{y}$. We assume that the prior distribution is a multivariate lognormal distribution $\ln\left(\mathbf{y}\right)\sim \mathcal{N}_d\left(\boldsymbol{\mu}_{log,prior},b\Sigma_{log,prior}\right)$. 
The mean $\boldsymbol{\mu}_{log,prior}$ and covariance matrix $\Sigma_{log,prior}$ are estimated from the samples $\{  \mathbf{y}^{*(i)}_{prior}  \}_{i=1}^{N_b}$, where $\mathbf{y}^{*(i)}_{prior}$ are the parameters estimated via ``coarse'' fitting to the field breakthrough curves from the TIERRAS dataset and $N_b$ is the number of breakthrough curves in the TIERRAS dataset (minus outliers, which are identified as explained below). For the power-law model, $\mathbf{y}^{*(i)}_{prior} = \mathbf{y}^{*(i)}_{Lap}$, while for the first-order exchange model, $\mathbf{y}^{*(i)}_{prior} = \mathbf{y}^{*(i)}_{Lap}$ if $\varepsilon^{(i)}_{RMSE}(v^{*(i)}_{Lap},\mathbf{y}^{*(i)}_{Lap})<\varepsilon^{(i)}_{RMSE}(v^{*(i)}_{mom},\mathbf{y}^{*(i)}_{mom})$ and $\mathbf{y}^{*(i)}_{prior} = \mathbf{y}^{*(i)}_{mom}$ otherwise, with $\varepsilon_{RMSE}$ defined in Eq \eqref{eq:RMSE}. The parameters $\mathbf{y}^{*(i)}_{Lap}$ and $\mathbf{y}^{*(i)}_{mom}$ are estimated using Laplace domain fitting (Section \ref{sec:methods-laplace-fit}) and matching of temporal moments (Section \ref{sec:methods-moments}), respectively. Temporal moments are not used in the power-law model because they are undefined. We treat parameters estimated from these two methods as ``coarse'' prior estimates because of their relatively low accuracy and high sensitivity to measurement errors. For the variance of the prior, we choose $b=2$ to prevent overfitting to the prior knowledge. 

The TIERRAS dataset analysis, using moment matching and Laplace fitting, reveals that the set of estimated parameters contains several outliers. To remove outliers, we iteratively compute $\boldsymbol{\mu}_{log,prior}$ and $\Sigma_{log,prior}$,  and at each iteration we remove any sample $i$ satisfying the condition $\norm{\Sigma_{log,prior}^{-1/2}\left(\mathbf{y}^{*(i)}_{prior}-\boldsymbol{\mu}_{log,prior}\right)}>4$, where  $\Sigma^{-1/2}_{log,Lap}$ is the symmetric matrix that satisfies $\Sigma^{-1/2}_{log,prior}\Sigma^{-1/2}_{log,prior}=\Sigma^{-1}_{log,prior}$. This condition defines outliers as extreme values with a probability of exceedance of $0.1\%$ (Chi-square distribution). It is worth noting that the outliers are not breakthrough curves that cannot be represented by the models; rather, they are cases where moment matching and Laplace fitting perform poorly due to high sensitivity to measurement errors, model mismatch (even if small), and low sampling frequency. Sets of parameters that differ too much from most other estimations are an indicator of unphysical estimations. These outliers are not removed elsewhere in this work.

The breakthrough curves of the synthetic dataset are calculated with a unit mass of the tracer $\hat{M_{0}}$. The synthetic dataset is obtained by solving Eq \eqref{eq:memory_ADE_dimensionless} for the samples $\{\mathbf{y}^{(i)} \}_{i=1}^{N_\text{synth}}$ from the prior distribution, yielding the ensemble of solutions $\{\hat{c}^{(i)}(\hat{t}) \}_{i=1}^{N_\text{synth}}$, where $N_{\text{synth}} =1000$ is the size of the set that we use in this work. We sample the numerical solution with $\Delta\hat{t}=1/150$ and $\hat{t}\in[0,24]$. The upper bound of 24 was selected as the smallest integer exceeding the maximum observed value of $t^*/t_{\text{peak}}^*$ found in the dataset, where $t^*$ are the sampling times. $t_{\text{peak}}^*$ is used as a surrogate for $L/v$ when constraining measured instead of synthetic breakthrough curves.

\begin{figure}[htb!]
	\centering
    \includegraphics[angle=0,width=\textwidth]{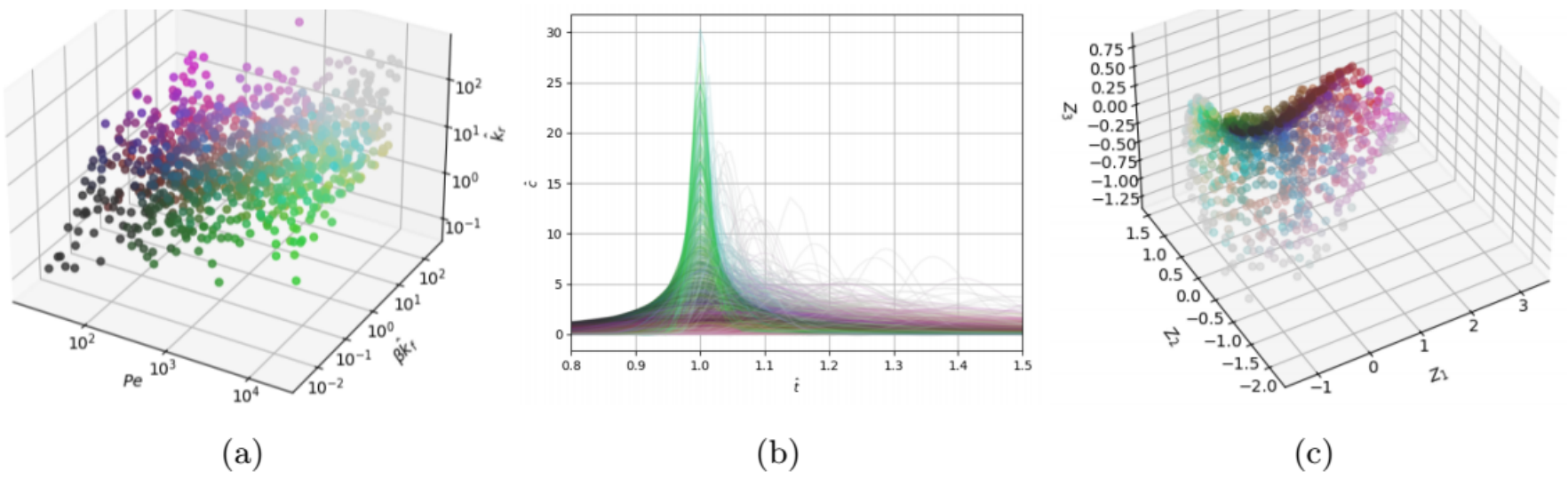}
	\caption{Representations of the synthetic dataset ($N_{\text{synth}} =1,000$) for the ADE model with a first-order memory function in (a) parameter space, (b) breakthrough curves form, and (c) embedded in the space of KL coefficients, only first three coefficients being represented. The same synthetic datapoint has matching colors in the three plots. The colorscale is created to represent different combinations of parameters. The visualization of the KL embedding of the synthetic dataset and its association to the parameters (particularly when combined with measurements, as shown in Figure \ref{fig:measured-dimensionless-and-embedding}c) provides a way to evaluate how well the chosen distribution of parameters matches the measured data and which parameters could be modified to improve it.}
	\label{fig:synthetic-construction-and-embedding}
\end{figure}

Figs \ref{fig:synthetic-construction-and-embedding}a and \ref{fig:synthetic-construction-and-embedding}b show the samples $\mathbf{y}$ and the corresponding synthetic breakthrough curves for the advection-dispersion model with a first-order memory function, respectively. 

\subsection{Reduced-Order Representation of the Synthetic Dataset} \label{sec:methods-KL-decomp}

If the samples of $\mathbf{y}$ are taken from any random distribution, then the breakthrough curve $\hat{c}(\hat{x}=1,\hat{t},\mathbf{y})$ is a continuous stochastic process (with covariance function $C_{\hat{c}}(\hat{t},\hat{t}')$) that can be approximated with the truncated KL expansion $\tilde{c}(\hat{t},\mathbf{y})$ as
\begin{linenomath*}\begin{equation}
    \hat{c}(\hat{x}=1,\hat{t},\mathbf{y}) \approx \tilde{c}(\hat{t},\mathbf{y}) = \bar{c}(\hat{t})+\sum_{j=1}^{N} \sqrt{\lambda_j} \eta_j(\mathbf{y}) \phi_j(\hat{t}) = \bar{c}(\hat{t})+\sum_{j=1}^{N}Z_j(\mathbf{y})  \phi_j(\hat{t}) = \bar{c}+ \mathbf{Z} \cdot\boldsymbol{\phi}
    \label{eq:KL}
\end{equation}\end{linenomath*}
where $\lambda_j$ are eigenvalues (arranged in decreasing order) and $\phi_j(\hat{t})$ are the corresponding eigenfunctions satisfying the orthogonality condition $\int_{\Omega_{\hat{t}}} \phi_i\phi_j d\hat{t} = I_{i,j}$ ($\Omega_{\hat{t}}$ is the time domain and $I$ is the identity matrix), $\eta_j$ are uncorrelated random variables with zero mean and unit variance, and $\mathbf{Z}$ are centrally distributed with a covariance matrix $\boldsymbol{\lambda}I$. 

The eigenfunctions and eigenvalues are found as the solution to the equation \cite{li_physics-informed_2022}
\begin{linenomath*}\begin{equation}
    \lambda_j \phi_j(\hat{t}) = \int_{\Omega_{\hat{t}}}  C_c(\hat{t},\hat{t}') \phi_j(\hat{t}') d\hat{t}'.
\end{equation}\end{linenomath*}
The number of terms $N$ is chosen so that the error $\hat{c}(x^*,\hat{t}) - \bar{c}(\hat{t})-\sum_{j=1}^{N}Z_j  \phi_j(\hat{t})$ is as low as desired. Here we use $N=20$, which results in a median RMSE (equation  \eqref{eq:RMSE}) of 0.0004\%, well below other errors associated to our problem. In this work, we compute $\bar{c}(\hat{t})$ and $C_c(\hat{t},\hat{t}')$ as sample statistics of the synthetic dataset  $\{\hat{c}^{(i)}(\hat{t}) \}_{i=1}^{N_\text{synth}}$.

An important property of the KL reduced-order model is the isometry or distance invariance of the transformation $\tilde{c}\to\mathbf{Z}$  \cite{jolliffe_mathematical_2002}. Consider the L2 distance between two dimensionless breakthrough curves:
\begin{linenomath*}\begin{equation}
    \norm{ \hat{c}_1 - \hat{c}_2} =\int_{\omega_{\hat{t}}} \left(\hat{c}_1 (x=1,\hat{t}) - \hat{c}_2(x=1,\hat{t})\right)^2 d\hat{t}\approx \norm{ \tilde{c}_1 - \tilde{c}_2}.
\end{equation}\end{linenomath*}
Replacing $c_1$ and $c_2$ with their KL expansions we obtain 
\begin{linenomath*}\begin{align}
    \norm{ \tilde{c}_1 - \tilde{c}_2} &= \int_{\omega_{\hat{t}}} \left(\bar{c}+ \mathbf{Z}_1\cdot \boldsymbol{\phi}- \bar{c}- \mathbf{Z}_2 \cdot\boldsymbol{\phi}\right)^2  d\hat{t} = \int_{\omega_{\hat{t}}} \left(\left(\mathbf{Z}_1 - \mathbf{Z}_2\right) \cdot\boldsymbol{\phi}  \right)^2  d\hat{t}   \\\nonumber
    &=  \int_{\omega_{\hat{t}}}\left(\mathbf{Z}_1 - \mathbf{Z}_2\right)^T \boldsymbol{\phi}\boldsymbol{\phi}^T \left(\mathbf{Z}_1 - \mathbf{Z}_2\right)   d\hat{t} = \left(\mathbf{Z}_1 - \mathbf{Z}_2\right)^T 
 \int_{\omega_{\hat{t}}}\boldsymbol{\phi}\boldsymbol{\phi}^T     d\hat{t} \left(\mathbf{Z}_1 - \mathbf{Z}_2\right).
\end{align}\end{linenomath*}
Noting that $\int_{\omega_{\hat{t}}}\boldsymbol{\phi}\boldsymbol{\phi}^T    d\hat{t} = I$ due to the orthogonality of $\boldsymbol{\phi}$, 
\begin{linenomath*}\begin{equation}
    \norm{ \tilde{c}_1 - \tilde{c}_2} = \left(\mathbf{Z}_1 - \mathbf{Z}_2\right)^T 
 I \left(\mathbf{Z}_1 - \mathbf{Z}_2\right) = \norm{ \mathbf{Z}_1 - \mathbf{Z}_2},
\end{equation}\end{linenomath*}
i.e., the L2 difference of functions is equal to the L2 norm of the difference of the KL coefficient vectors. 

The synthetic dataset for the advection-dispersion model with a first-order memory function embedded in the space of the first three KL coefficients is shown in Figure \ref{fig:synthetic-construction-and-embedding}c.

 \subsection{Estimation of KL Coefficients for Measured Breakthrough Curves}\label{sec:methods-Z-MAP}

To estimate parameters, we need to invert the function $\mathbf{y}\to\mathbf{Z}$, but first we need an estimate $\mathbf{Z}^*$ of the KL coefficients for each measured breakthrough curve. There are two difficulties with this. First, the dimensionless form of a measured breakthrough curve depends on the advection velocity $v$, which is unknown. Second, the sampling times of measured breakthrough curves differ from and are typically sparser than those in the synthetic dataset. To address the first, we treat the estimated KL coefficients as a function $v$, denoted $\mathbf{Z}^*(v)$. To address the second challenge, we use a Maximum a Posteriori (MAP) estimate of the KL coefficients from sparse measurements.

The MAP estimate of the KL coefficients is obtained by maximizing the Bayesian posterior distribution of $\mathbf{Z}^*(v)$ defined as
\begin{linenomath*}\begin{equation}
\mathbf{Z}^*(v)=\max_{\mathbf{Z}(v)} p(\mathbf{Z}(v)|c^*) = 
\max_{\mathbf{Z}(v)} [p(c^*|\mathbf{Z}(v))p(\mathbf{Z}(v))]
\label{eq:MAP_Zv}
\end{equation}\end{linenomath*}
where $p(\mathbf{Z}(v)|c^*)$ is the posterior distribution of the parameters given the breakthrough curve concentration measurements $c^*$, $p(c^*|\mathbf{Z}(v))$ is the likelihood, and $p(\mathbf{Z}(v)))$ is the prior of $\mathbf{Z}$.

The likelihood function is a model of the mismatch between the KL model prediction of the breakthrough concentration and the measured breakthrough curve. We assume that the mismatch values are normally distributed, i.e., 
\begin{linenomath*}\begin{equation*}
    p(c^*|\mathbf{Z} (v))= \frac{1}{\sqrt{2\pi\sigma_c^2}}\exp\left(-\frac{\sum_{i=1}^{N^*}\left(\frac{c^*_i \Delta^*_i}{\sum_{j=1}^{N^*} c^*_j\Delta^*_j} - \frac{\left(\overline{c}^{(I)}\left(\frac{t^*_i v}{L}\right) + \mathbf{Z}\cdot\boldsymbol{\phi}^{(I)}\left(\frac{t^*_iv}{L}\right)\right)\Delta^*_i}{\sum_{j=1}^{N^*} \overline{c}^{(I)}\left(\frac{t^*_jv}{L}\right)\Delta^*_j}\right)^2}{N^* 2\sigma_c^2}\right),
\end{equation*}\end{linenomath*}
with zero mean and variance $\sigma_c^2$. Here, $N^*$ is the number of measurements in the given breakthrough curve, $c_i^*$ are the measured concentrations at times $t^*_i$,  and $\Delta^*_i$ are the corresponding time intervals. In the field, breakthrough curves, only non-zero concentration values are reported. To improve MAP estimates, we add zero concentration values outside of the time interval where measurements are reported, with a sampling frequency equal to the lowest shown frequency.   The superscript $(I)$ symbolizes that the function $\boldsymbol{\phi}^{(I)}$ is linearly interpolated from the synthetic dataset. This interpolation introduces minimal error due to the high resolution of the synthetic solutions.

Based on the definition of the KL expansion in Eq \eqref{eq:KL}, the prior distribution of $\mathbf{Z}$ is multivariate Gaussian $\mathbf{Z}\sim \mathcal{N}_N (\mathbf{0}, \boldsymbol{\lambda} I) $. Noting that the maximum of the posterior distribution is equal to the minimum of the natural logarithm of the posterior, we can rewrite Eq \eqref{eq:MAP_Zv} as 
\begin{linenomath*}\begin{equation}
\mathbf{Z}^*(v) = \min_{\mathbf{Z}} 
\left( \frac{\sum_{i=1}^{N^*}\left(\frac{c^*_i \Delta^*_i}{\sum_{j=1}^{N^*} c^*_j\Delta^*_j} - \frac{\left(\overline{c}^{(I)}\left(\frac{t^*_i v}{L}\right) + \mathbf{Z}\cdot\boldsymbol{\phi}^{(I)}\left(\frac{t^*_iv}{L}\right)\right)\Delta^*_i}{\sum_{j=1}^{N^*} \overline{c}^{(I)}\left(\frac{t^*_jv}{L}\right)\Delta^*_j}\right)^2}{N^* \sigma_c^2} 
+\mathbf{Z}^T (\boldsymbol{\lambda}I)^{-1} \mathbf{Z} \right).
\label{eq:MAP_Zv_1}
\end{equation}\end{linenomath*}
This is a linear least squares problem, which has a unique solution and is relatively fast to solve. The optimal value of the hyperparameter $\sigma_c$  is selected to minimize the expected L2 validation error. For this, each breakthrough curve is randomly divided into 80\% training and 20\% validation samples. We use the geometric mean of validation errors across all breakthrough curves to account for variability in their magnitudes. Once the optimal value of $\sigma_c$ is obtained, we estimate $\mathbf{Z}^*(v)$ for each breakthrough curve without removing any measurements.

This process allows us to obtain the most likely upsampling of the measured breakthrough curves onto the high-frequency sampling of the synthetic dataset. If interpolation is applied between measurements, the estimated parameters may become highly sensitive to the selected interpolation method. This sensitivity increases the risk of overfitting the KL coefficients. The truncation of the KL expansion reduces the number of linear interpolations and the size of the linear least-squares problem in Eq \eqref{eq:MAP_Zv_1}.

\begin{figure}[htb!]
	\centering
    \includegraphics[angle=0,width=\textwidth]{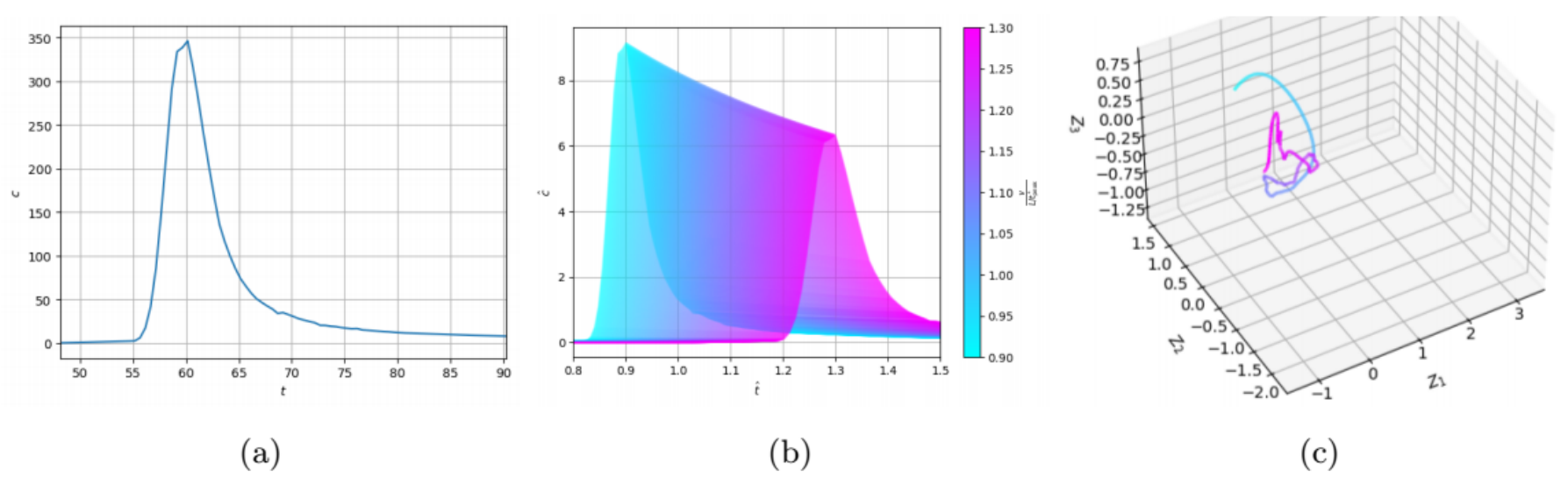}
	\caption{KL embedding process of a measured breakthrough curve. (a) Measured breakthrough curve. (b) Different dimensionless breakthrough curves corresponding to the same dimensional measurement but different advection velocities. (c) Embedding of the dimensionless breakthrough curves corresponding to a single measured breakthrough curve in the space of KL coefficients, only first three coefficients being represented. The definition of the KL space is the same as in Figure \ref{fig:synthetic-construction-and-embedding}c.}
	\label{fig:measured-dimensionless-and-embedding}
\end{figure}

Figure \ref{fig:measured-dimensionless-and-embedding} shows the process of embedding a measured breakthrough curve (depicted in Figure \ref{fig:measured-dimensionless-and-embedding}a) in the KL space of the synthetic dataset. Figure \ref{fig:measured-dimensionless-and-embedding}b illustrates that for the same dimensional breakthrough curve, different values of $v$ correspond to different dimensionless breakthrough curves. Figure \ref{fig:measured-dimensionless-and-embedding}c demonstrates that these dimensionless breakthrough curves form a line in the KL space, with each point on the line corresponding to a different value of $v$. This provides a dimensionless KL embedding of the breakthrough curve, before we have estimated the velocity.

\subsection{Projected Barycentric Interpolation}\label{sec:methods-PBI}

In this section, we develop a map from $\mathbf{Z}^*(v)$ to $\mathbf{y}^*$, the parameter vector corresponding to the field breakthrough curve described by the KL coefficient vector  $\mathbf{Z}^*(v)$. Since the dimensionality of $\mathbf{y}^*$ (here, $d=3$) is smaller than that of $\mathbf{Z}^*(v)$, such a map is unique.  The mapping is done by first projecting $\mathbf{Z}^*(v)$ on the manifold defined by $\mathbf{Z}$ corresponding to the synthetic dataset (this manifold is bounded for bounded parameters and $d$-dimensional), and then estimating $\mathbf{y}^*$ as the coordinate vector of the point on the manifold. The isometry of the KL transformation makes it so that a projection that minimizes the Euclidean distance to the manifold provides a least-squares fit of the measured breakthrough curve. Because of model error (model misspecification) and measurement errors, the Euclidean distance to the manifold is greater than zero, since the considered transport models do not describe the breakthrough curves exactly. For a given measured breakthrough curve, the Euclidean distance to the manifold depends on $v$. 

To estimate the value of the advection velocity $v^*$, we look for the value of $\mathbf{Z}^*(v)$ (Eq \eqref{eq:MAP_Zv}) that has the minimum distance to its projection onto the manifold \cite{park2025robustuniversalinferencemisspecified}, thus minimizing the L2 difference between the field and synthetic breakthrough curves. To project a measurement point $\mathbf{Z}^*(v)$ onto the manifold, we locally approximate the manifold as the simplex formed by the $d+1$ nearest neighbors to $\mathbf{Z}^*$.  The $d+1$ points on the manifold create a local triangulation since $d$, the dimensionality of $y$, is also the topological dimension of the manifold. Then, we project the measurement onto the linear subspace defined by the simplex. The projected point is
\begin{linenomath*}\begin{equation}
    \mathbf{r}^*(v) = \mathbf{r}_1 + \mathrm{proj}_{\mathbf{r}_2-\mathbf{r}_1} \mathbf{Z}^* (v) + \dots + \mathrm{proj}_{\mathbf{r}_{d+1}-\mathbf{r}_d} \mathbf{Z}^*(v),
\end{equation}\end{linenomath*}
where $\mathbf{r}_i, i=1,2,\dots,d+1$ are the vertices of the simplex. If the projected point falls outside of the simplex (if any barycentric coordinate is negative), we remove from the simplex its vertex that is furthest from the measurements. We iterate until the projection falls inside the simplex or until only one vertex (the nearest neighbor) is left. The barycentric coordinates vector $\boldsymbol{\rho}$ is a $d'$ dimensional vector $\boldsymbol{\rho} = \mathbf{T}^{-1} \left(\mathbf{r}-\mathbf{r}_{d'+1}\right)$, where $d'+1$ is the number of leftover vertices in the simplex. The last barycentric coordinate is $\rho_{d'+1}=1-\sum_{i=1}^{d'} \rho_i$. $\mathbf{T}^{-1}$ is the inverse of a square matrix $\mathbf{T}$ formed by the vector edges of the simplex that share the vertex $d'+1$, such that $\mathbf{T}=[\mathbf{r}_1-\mathbf{r}_{d'+1},\mathbf{r}_2-\mathbf{r}_{d'+1},\dots,\mathbf{r}_{d'}-\mathbf{r}_{d'+1}]$.

\begin{figure}[htb!]
	\centering
    \includegraphics[angle=0,width=\textwidth]{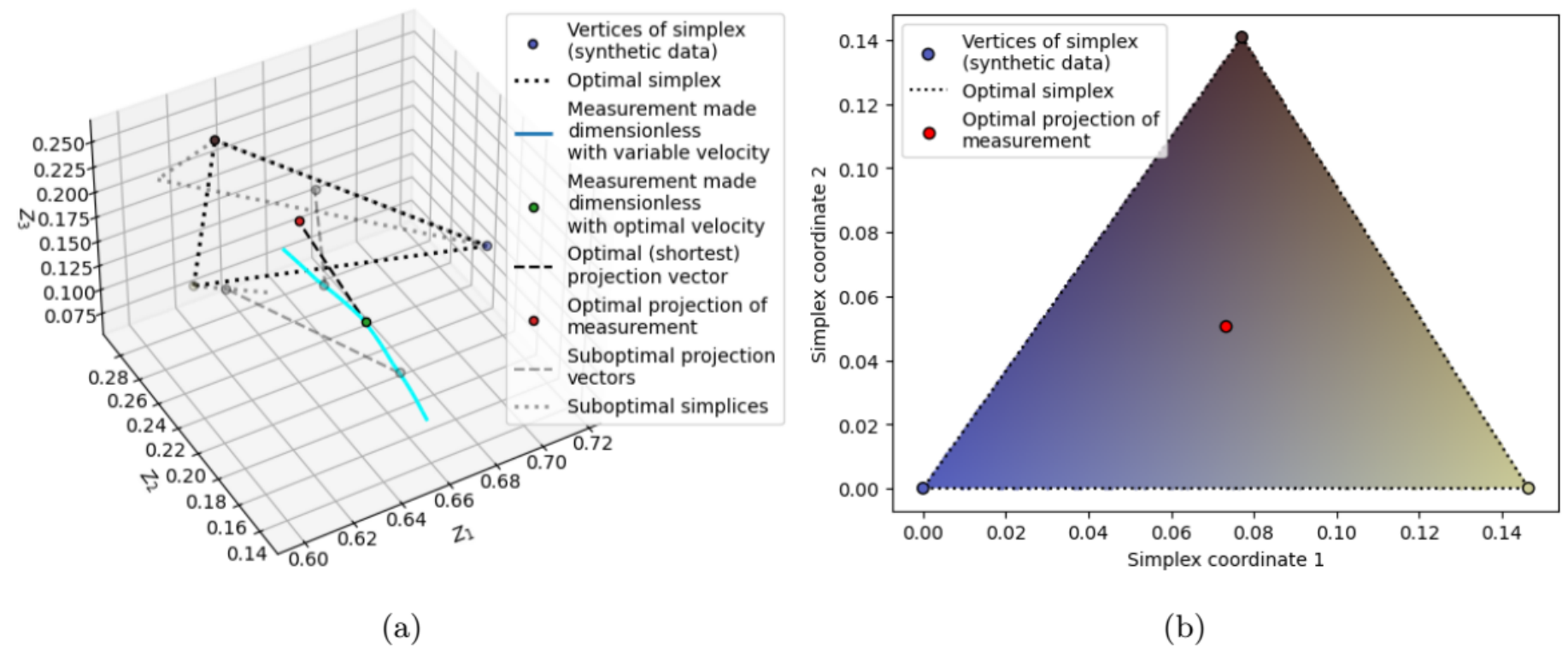}
	\caption{Illustration of the PBI method for the example shown in Figures \ref{fig:synthetic-construction-and-embedding} and \ref{fig:measured-dimensionless-and-embedding}, but with rescaled color scale: (a) KL embedding of synthetic data forming the vertices of the simplex and different dimensionless forms of the field breakthrough curve corresponding to different advection velocities. Optimal and suboptimal projections with their corresponding simplices, and (b) Local coordinate representation of a simplex with the color scale corresponding to parameters interpolated by barycentric interpolation.}
	\label{fig:PBI}
\end{figure}

The projection vector (sometimes called rejection vector) $\mathbf{Z}^*(v)-\mathbf{r}^*(v)$ shows the difference between the measurement and the projected point on the manifold, and its magnitude $\norm{ \mathbf{Z}^*(v)-\mathbf{r}^*(v)}$ is an estimate of model mismatch and measurement errors. We obtain an estimate of the velocity as
\begin{linenomath*}\begin{equation}\label{eq:vel_opt}
    v^*=\min_{v}\norm{ \mathbf{Z}^*(v)-\mathbf{r}^*(v)}.
\end{equation}\end{linenomath*}
This optimization is highly nonlinear and has many local minima. For this reason, we obtain the global minimum by exhaustive search in the range (from 0.9 to 1.5 times the peak velocity) with a resolution of 0.5\% of the peak velocity. Figure \ref{fig:PBI}a shows the projection process for an example measurement. Possible dimensionless forms (each corresponding to a different advection velocity) of a field breakthrough curve form a one-dimensional curve in the KL space. Each point on this one-dimensional curve is projected onto a simplex of the synthetic dataset. Since the projection length is an indicator of the magnitude of errors, its minimum defines the optimal velocity $v^*$ and the optimal projected measurement $\mathbf{r}^*(v^*)$.

Finally, the parameters estimated by PBI are $\ln\left(\mathbf{y}^*\right) = \sum_{k=1}^{d'+1} \ln\left(\mathbf{y}^{(k)}\right) \rho_k$, where $\mathbf{y}^{(k)}$ are the parameters used to create the synthetic data points corresponding to each of the vertices of the optimal simplex, and $\rho_k$ are the barycentric coordinates of the optimal projection $\mathbf{r}^*(v^*)$. The barycentric interpolation process is represented in Figure \ref{fig:PBI}a. The color at each of the vertices of the simplex represents the set of parameters $\mathbf{y}$ used to create the corresponding breakthrough curve in the synthetic dataset. The colors of the vertices are barycentrically interpolated inside of the simplex creating a color gradient, corresponding to estimated values of the parameters in its interior.

\section{Other Methods for Parameter Estimation} \label{sec:other-methods}

In this section, we introduce variations on the DSTE framework and state-of-the-art methods for estimating parameters, which we use to obtain the prior distribution of parameters for DSTE, to benchmark, and to further refine estimates from DSTE-KL-PBI.

\subsection{Alternative Methods Within the DSTE Framework}

We propose alternative methods within the DSTE framework that use the same dimensionally reduced training set but vary in how the manifold is learned and interpolated.

\subsubsection{DSTE-KL Nearest-Neighbor Interpolation (NNI)}

The Nearest-Neighbor Interpolation (NNI) of the reduced space of a field breakthrough curve on the training set is defined as $\mathbf{y}^*=\mathbf{y}^{(i^*)}$, where 
\begin{linenomath*}\begin{equation*}
    i^*,v^* = \min_{i,v}\norm{\mathbf{Z}^*(v)-\mathbf{Z}^{(i)}}.
\end{equation*}\end{linenomath*}
For the considered dataset, the exhaustive search for the minimum is computationally easy because the cost of calculating distances in the KL coefficient space is significantly smaller than the cost of obtaining forward solutions. No projection is required for this algorithm, as the nearest neighbor is well-defined regardless of whether the point is inside or outside the manifold.  This method is almost exactly equivalent to an exhaustive search of the dimensionless synthetic breakthrough curve that best matches the measured breakthrough curve in the L2 sense, which solution is given by the minimization problem 
\begin{linenomath*}\begin{equation}
v^*,\mathbf{y}^* = \min_{v,\mathbf{y}}\sum_{i=1}^{N^*} \left(\frac{c^*_i \Delta^*_i}{\sum_{j=1}^{N^*} c^*_j \Delta^*_j}-\frac{\hat{c}^{(I)}\left(\hat{x}=1,\frac{t^*_iv}{L},\mathbf{y}\right)\Delta^*_i}{\sum_{j=1}^{N^*}\hat{c}^{(I)}\left(\hat{x}=1,\frac{t^*_jv}{L},\mathbf{y}\right)\Delta^*_j}\right)^2.
\label{eq:exhaustive_search}
\end{equation}\end{linenomath*}
We note that Eq \eqref{eq:exhaustive_search} constitutes a non-KL version of DSTE. The main advantage of the NNI method over the exhaustive search method is the computational efficiency of only having to interpolate $N$ times instead of $N_b$ times per value of $v$ at the cost of small errors associated with the truncation of the KL expansion.

\subsubsection{DSTE-KL Forward Deep Neural Networks (FDNN)}

We learn the relationship between $\mathbf{Z}$ and $\mathbf{y}$ using a deep neural network (DNN) trained in the forward direction, $\mathbf{Z}(\bar{\mathbf{y}}^{(i)}) \approx NN(\bar{\mathbf{y}}^{(i)},\boldsymbol{\theta})$, where $\boldsymbol{\theta}$ are the DNN parameters and $\bar{\mathbf{y}}^{(i)}= \Sigma^{-1/2}_{log,prior}\left(\mathbf{y}^{(i)}-\boldsymbol{\mu}_{log,prior}\right)$ are the normalized parameters.

We use a feed-forward fully connected neural network with three hidden layers of 50 neurons each, with ReLU activation functions. We find this architecture by manually fine-tuning hyperparameters to optimize the validation errors of a 90-10 training-validation split. We test the results with 100 additional test runs. The parameters of the neural networks are obtained as
\begin{linenomath*}\begin{equation}\label{eq:DNNloss_latent}
\boldsymbol\theta^* = 
\sum_{i=1}^{N_\text{synth}}
\min_{\boldsymbol\theta} 
\norm{\mathbf{Z}^{(i)} - \mathcal{NN}\left(\bar{\mathbf{y}}^{(i)},\boldsymbol{\theta}\right)}^2,
\end{equation}\end{linenomath*}
which we solve using the Adam optimizer \cite{kingma_adam_2017} with 600 training steps (when the testing error reaches a minimum), and training by batches of 100 samples each, of the synthetic dataset $\{\bar{\mathbf{y}}^{(i)},\mathbf{Z}^{(i)} \}_{i=1}^{N_\text{synth}}$.

Last, we obtain the estimate of $\mathbf{y}$ as the solution of the following optimization problem, 
\begin{linenomath*}\begin{equation}
v^*,\bar{\mathbf{y}}^* = \min_{v,\bar{\mathbf{y}}}
\sum_{i=1}^{N^*}\left(\frac{c^*_i\Delta^*_i}{\sum_{j=1}^{N^*} c^*_j\Delta^*_j} - \frac{\left(\overline{c}^{(I)}\left(\frac{t^*_i v}{L}\right) + \mathcal{NN}\left(\bar{\mathbf{y}},\boldsymbol{\theta}\right)\cdot\boldsymbol{\phi}^{(I)}\left(\frac{t^*_iv}{L}\right)\right)\Delta^*_i}{\sum_{j=1}^{N^*} \overline{c}^{(I)}\left(\frac{t^*_jv}{L}\right)\Delta^*_j}\right)^2
\label{eq:KL-NN-inversion}
\end{equation}\end{linenomath*}
which we also minimize using the Adam optimizer. To be optimized with this method, the interpolation method was programmed to be compatible with automatic differentiation. To initialize the optimizer we use parameters obtained with DSTE-KL-PBI.

Since priors (from Laplace domain fitting and moment matching) are significantly worse than the results of the Neural Network, it was not worth trying to obtain a MAP estimate instead of a Maximum Likelihood estimate (as obtained in Eq \eqref{eq:KL-NN-inversion}, assuming a Gaussian distribution of errors), given the trade-off of having to fit a regularization weight.

\subsubsection{DSTE-KL Inverse Deep Neural Networks (IDNN)}\label{sec:methods-inverse}

In this method, a DNN is trained to directly map KL coefficients $\mathbf{Z}$ to the parameters $\mathbf{y}$. A similar inverse DNN approach for mapping measurements to the system's parameters was used in \cite{vihrs_using_2022, lenzi_neural_2023}. We formulate this approach in the space of KL coefficient $\mathbf{Z}$ of the breakthrough curve, i.e., we train the "inverse'' DNN $\bar{\mathbf{y}} \approx \mathcal{NN}_{inv}(\mathbf{Z},\boldsymbol{\theta})$ using the same labeled dataset $\{\bar{\mathbf{y}}^{(i)},\mathbf{Z}^{(i)} \}_{i=1}^{N_\text{synth}}$ that we use to train the forward KL-DNN surrogate model and the same normalization. 

The parameters in the inverse DNN are found by solving the minimization problem 
\begin{linenomath*}\begin{equation*}
    \boldsymbol\theta^*_{inv} = 
\sum_{i=1}^{N_\text{synth}}
\min_{\boldsymbol\theta} 
\norm{\bar{\mathbf{y}}^{(i)}  - \mathcal{NN}_{inv}(\mathbf{Z}^{(i)},\boldsymbol{\theta})}^2.
\end{equation*}\end{linenomath*}
We use the same neural network architecture and optimization process as for the forward problem. Then, the parameter estimate is given as $\bar{\mathbf{y}}^* =\mathcal{NN}_{inv}(\mathbf{Z}^*(v^*),\boldsymbol{\theta}) $, using the same estimate of $\mathbf{Z}^*(v^*)$ that we obtained for DSTE-KL-PBI (Eqs. \eqref{eq:MAP_Zv_1} and \eqref{eq:vel_opt}).

The inverse Neural Network is compelled to extrapolate when its output is influenced by field breakthrough curves—or the associated KL coefficients $\mathbf{Z}^*(v^*)$—that lie beyond the boundaries of the training set manifold. To avoid extrapolation, we extend our synthetic data by randomly perturbing $\mathbf{Z}^{(i)}$ in the training set (while $\mathbf{y}^{(i)}$ are kept unperturbed).  This method is known as data augmentation \cite{cromwell_estimating_2021}, and similar methods have been used in the past to address model misspecification \cite{kennedy_bayesian_2001}. The primary challenge with the method is determining the magnitude and quality of the perturbations that most effectively augment the training set. The more complex the input, the harder this is to do. As mentioned in Section \ref{sec:methods-PBI}, our reduced dimension manifold projection can give a good estimate of model misspecification and measurement errors by taking the distance from the data point to its projection on the synthetic data manifold. We utilize this to our advantage, restricting the direction of augmentation. We define the synthetic points for augmentation as $\{\bar{\mathbf{y}}_{\text{aug}}^{(i)},\mathbf{Z}_{\text{aug}}^{(i)} \}_{i=1}^{N_\text{synth}\times N_{\text{aug}}}$, where $\mathbf{Z}_{\text{aug}}^{(i)}=\mathbf{Z}^{\left\lceil\frac{i}{N_{\text{aug}}}\right\rceil}+a\mathbf{X}$ and $\bar{\mathbf{y}}_{\text{aug}}^{(i)}=\bar{\mathbf{y}}^{\left\lceil\frac{i}{N_{\text{aug}}}\right\rceil}$. $\mathbf{X}\sim\mathcal{N}(\boldsymbol{\mu}_{\mathbf{r}},\boldsymbol{\Sigma}_{\mathbf{r}})$, where $\boldsymbol{\mu}_{\mathbf{r}}$ and $\boldsymbol{\Sigma}_{\mathbf{r}}$ are the mean and covariance matrix of the set of projection vectors $\{  \mathbf{r}^{*(i)} (v^{*(i)}) \}_{i=1}^{N_b}$. 

For the inverse DNN, we use 10 terms in the KL expansion instead of 20 terms. This increases the RMSE of the KL expansion to 0.0044\%. This is still a relatively low error, but it decreases the number of components in which augmentation is needed, improving the final error for the same number of augmentation points. We use $N_{aug}=30$ and $a=0.5$.

\subsection{State of the Art Methods for Parameter Estimation}\label{sec:methods-coarse}

Here, we consider three methods for parameter estimation. The Laplace domain fitting and moment matching are computationally efficient but less accurate. We use these methods to sample the prior distribution of the parameters. The global random optimization method, combined with a numerical solver, is the standard approach that incurs high computational costs, and we utilize it to benchmark the proposed parameter estimation methods.  

\subsubsection{Laplace Domain Fitting}\label{sec:methods-laplace-fit}

To estimate $ \mathbf{y}^{*(i)}_{Lap}$, we first define a normalized form of Laplace-transformed theoretical breakthrough curve $ f(\hat{s},\mathbf{y})  = \ln\left( \frac{\hat{C}(\hat{x}^*=1,\hat{s},\mathbf{y}, \hat{M}_0)}{\hat{C}(\hat{x}^*=1,0,\mathbf{y},\hat{M}_0)}  \right)$, which is independent of the mass released $\hat{M}_0$. We can write an analytical expression for $f(\hat{s},\mathbf{y})$ as 
\begin{linenomath*}\begin{equation} \label{eq:laplace-fitting}
    f(\hat{s},\mathbf{y}) = \frac{1}{2}\left(Pe-\sqrt{Pe\left(4\left(\hat{s} +  \beta \hat{s}\hat{G}(\hat{s},\mathbf{y})\right)+Pe\right)}\right) +\ln\left(B(\hat{s},\mathbf{y})\right).
\end{equation}\end{linenomath*}
Then, we can estimate $\mathbf{y}$ in Laplace domain for a breakthrough curve as
\begin{linenomath*}\begin{equation}
    v^*_{Lap},\mathbf{y}^*_{Lap} = \min_{v,\mathbf{y}}  \norm{ f\left(\frac{sx}{v},\mathbf{y}\right) - f^*(s)}^2,
\end{equation}\end{linenomath*}
where $f^*(s)= \ln\left( \frac{C^*(s)}{C^*(0)}  \right)$, and $C^*(s)$ is the Laplace transform of the measured breakthrough curve $c^*(t)$ obtained by numerical integration. We employ \textit{least\_squares} routine in \textit{scipy.optimize} \cite{2020SciPy-NMeth} to solve this minimization problem. It uses the Trust Region Reflective method \cite{branch_subspace_1999}. A similar approach was presented in  \citeA{seo_moment-based_2001}.

\subsubsection{Moments Matching}\label{sec:methods-moments}

The temporal moments of the breakthrough curves were used to estimate parameters of river transport models in \citeA{seo_moment-based_2001} and \citeA{gonzalez-pinzon_scaling_2013}. Moment matching consists of fitting the analytical expressions of the first $n$ moments to the first $n$ moments computed from the field breakthrough curve. We use the first four central moments, where the $k$-th central temporal moment of a breakthrough curve is defined as $m_k(x) = \frac{\int_0^\infty t^k c(x,t) dt}{\int_0^\infty c(x,t) dt}$ for $k=1$ and $m_k(x)=\frac{\int_0^\infty (t-m_1(x))^k c(x,t) dt}{\int_0^\infty c(x,t) dt}$  for $k>1$. The matching can be done in several ways. Here, we use a least square approach
\begin{linenomath*}\begin{equation}\label{eq:moments-loss}
    v^*_{mom},\mathbf{y}^*_{mom} = \min_{\mathbf{y}} \left(\sum_{k=1}^{4} \left(m_k(x^*,\mathbf{y})^{1/k}-{m_k^*}^{1/k}\right)^2\right),
\end{equation}\end{linenomath*}
where $m_k$ are the analytical values of the central moments and $m_k^*$ are the measured central moments at $x^*$. The exponent $\frac{1}{k}$ is used to make different moments dimensionally equivalent. The minimum is found by the \textit{least\_squares} routine in \textit{scipy.optimize}.

The form of the first four analytical temporal moments for a generic memory function was obtained by \citeA{aghababaei_temporal_2023} for a semi-infinite domain without upstream dispersion (Column 1 in Table \ref{tab:problems-comparison}), where $\mu_k(0)$ is known. Here, we extend this approach to the initial and boundary conditions listed in Columns 2-4 of Table \ref{tab:problems-comparison}. In \cite{aghababaei_temporal_2023}, the moments were derived from the moment equations, which are obtained by applying the moment operator $\int_0^{\infty} \tau^k [\circ] d\tau$ to the differential equations. When the boundary condition is of the form $c(0,t)=f(t)$, $\mu_k(0)$ can be calculated directly from its definition.  For problems defined in a semi-infinite domain $\Omega_x = [0,\infty)$, where the upstream boundary condition is of the form $v c(x=0)-n_c D\frac{dc(x=0)}{dx}=M_0 \delta(t)$, the boundary conditions of the moment equations are $v\mu_0(0)-n_c D\frac{d\mu_0(0)}{dx}=M_0$ and $v\mu_k(0)=n_c D\frac{d\mu_k(0)}{dx}$  for $k\geq 1$, also satisfying that $\mu_k(\infty)$ must not grow exponentially fast. This form is applicable to all the problems in Table \ref{tab:problems-comparison}, with $n_c=0$, $1$, or $2$ for the problems with a semi-infinite domain without and with upstream dispersion and the one equivalent to an infinite domain (columns 1,2, and 4 of Table \ref{tab:problems-comparison}), respectively.

As shown by \citeA {aghababaei_temporal_2023}, the solution to the moments is of the form  $\mu_k(x) = \sum_{i=1}^{k} a_{k,i} x^i + \mu_k(0)$, where $a_{k,i}$ for $k$ up to 4 are found in Appendix A of \cite{aghababaei_temporal_2023} for a generic memory function. In general, all moments take this form, as can be seen by solving the governing equation for increasing moments iteratively and applying the condition that $\mu_k(x)$ must not grow exponentially fast as $x\to\infty$. After applying the aforementioned boundary conditions we obtain 
\begin{linenomath*}\begin{equation}
    \mu_0(0) = \frac{M_0}{v};\quad
    \mu_k(x) = a_{k,1} \left(x+n_c \frac{D}{v}\right) + \sum_{i=2}^{k} a_{k,i} x^i.  
\end{equation}\end{linenomath*}

The expressions for the temporal moments for the particular case of the Transient Storage Model in an infinite domain with equal forward and reverse rates was also obtained by \citeA{seo_moment-based_2001}. They apply a different method that consists of using the Laplace domain solution as a moment-generating function. Their results are consistent with ours.

\subsubsection{Global Random Optimization Method (LIPO)}

To refine results and use them as a benchmark method, we solve the problem using the numerical solver described in Section \ref{sec:methods-laplace-forward} and the stochastic global optimization method LIPO \cite{malherbe_global_2017}.  The LIPO method is a random search algorithm that discards regions for search if, given previous function evaluations, the Lipschitz parameter indicates that an optimum cannot exist in that region. Since the Lipschitz constant is not known a priori, the algorithm alternates between evaluating the function at different points and estimating the Lipschitz constant using all previous measurements. We use the Python implementation by \citeA{lipo_2021}, which combines the LIPO algorithm with a trust region algorithm to get better performance in the close vicinity of a minimum. The estimated parameters for this method are defined as $v^*,\mathbf{y}^*=\min_{v,\mathbf{y}}\varepsilon_{RMSE} (v,\mathbf{y})$, where $\varepsilon_{RMSE} (v,\mathbf{y})$ is the Root Mean Square Error of the breakthrough curve reconstructed with the numerical solver, as defined in Eq \eqref{eq:RMSE}. We use $\mathbf{y}^{*(i)}_{prior}$ as the initial guess for each breakthrough curve. The algorithm is constrained to maximum and minimum values of the parameters. We use $|\mathbf{y}^{*(i)}_{prior}-\boldsymbol{\mu}_{log,prior}|<4\cdot\mathrm{diag}\left(\Sigma_{log,prior}^{-1/2}\right)$ (where $\mathrm{diag}()$ symbolizes the diagonal of a matrix), for the parameter search space to be consistent with the synthetic dataset.

We also use the parameters estimated with DSTE-KL-PBI as the initial guess for the LIPO method. We find that this approach provides the most accurate estimates of the parameters. The LIPO's capacity to provide refinements stems from both the trust region steps of the algorithm and the LIPO steps. The LIPO steps can refine already good results because we use narrower constraints for refinement, reducing the search space, and because the better the initial guess, the larger the first discarded region in the method.

The loss of the optimization problem using LIPO scales as $\sim \left(\frac{V(\chi)}{N_{\text{ev}}}\right)^{\frac{1}{d+1}}$ \cite{malherbe_global_2017}, where $V(\chi)$ is the volume of the parameter searching set (using the log of the parameter because they span multiple orders of magnitude).

\subsection{Estimation of Parameters in the Advection-Dispersion Model Without Mass Exchange}\label{sec:ADE-fit}

For a semi-infinite domain without upstream dispersion or infinite domain (columns 1 and 3 respectively in Table \ref{tab:problems-comparison}), ADE, i.e., Eq \eqref{eq:memory_ADE_dimensional} with $\beta = 0$, has a simple analytical solution shown in Table \ref{tab:problems-comparison}. In this work, we only use the solution in the infinite domain. Because this solution is very fast to evaluate and has only two parameters, a least-squares estimate of the parameters can be easily found by solving the minimization problem
\begin{linenomath*}\begin{equation}
v^*, Pe^* = \min_{v,Pe}
\sum_{i=1}^{N^*}\left(\frac{c^*_i \Delta^*_i}{\sum_{j=1}^{N^*} c^*_j\Delta^*_j} - \frac{\hat{c}_{ADE}(\hat{x}=1,\frac{t^*_i v}{L},\hat{M}_0=1,Pe)\Delta^*_i}{\sum_{j=1}^{N^*} \hat{c}_{ADE}(\hat{x}=1,\frac{t^*_j v}{L},\hat{M}_0=1,Pe)\Delta^*_j}\right)^2,
\label{eq:optimization_ADE}
\end{equation}\end{linenomath*}
where $\hat{c}_{ADE}(\hat{x},\hat{t},\hat{M}_0,Pe)$ is the ADE solution shown in the last row of Table \ref{tab:problems-comparison}. We use the \textit{scipy} function \textit{optimize.least\_squares} to solve this least-squares problem.

Additionally, we fit the analytical ADE solution to the peak of the breakthrough curve by matching the analytical and the measured values of the time of the peak and the ratio between the second derivative of the concentration and the concentration at the peak. These methods give estimates of the transport parameters that ignore the effects of the breakthrough tails. Matching the times of the peak arrival leads to the explicit equation for the velocity $v = \frac{\sqrt{L^2-2 t^*_{\text{peak}}D}}{t^*_{\text{peak}}}$ and matching the ratios yields the implicit equation for the dispersion coefficient:
\begin{linenomath*}\begin{equation*}
    D = \frac{b (b^2 - 3) v^3 L - \sqrt{ 4 b^4 (b^2 - 1)^2 \frac{\frac{\partial^2 c}{\partial t^2}(t=t^*_{\text{peak}})}{c(t=t^*_{\text{peak}})}  v^4 L^4 - 2 b^2 (b^4 - 3) v^6 L^2}}{2b^2 \left(4\frac{\frac{\partial^2 c}{\partial t^2}(t=t^*_{\text{peak}})}{c(t=t^*_{\text{peak}})} b^2 L^2 - 3 v^2\right) },
\end{equation*}\end{linenomath*}
where $\alpha =\sqrt{\frac{D^2}{v^2}+L^2} - \frac{D}{v}$. The system of these two equations is solved by fixed-point iteration. The derivatives of the measured breakthrough curves are calculated with a second-order central scheme. 

\section{Results: Parameter Estimation in Transport Models From Field Breakthrough Curves}\label{sec:results}

In this section, we study the accuracy of the DSTE-KL-PBI method for estimating parameters of the advection-dispersion model with immobile exchange, described by first-order and power-law memory functions, using field breakthrough curve data. We compare DSTE-KL-PBI with other methods introduced in Section \ref{sec:other-methods}. We present results for Problem Formulation 4 with the initial and boundary conditions listed in the last column of Table \ref{tab:problems-comparison}. Results obtained for other initial boundary conditions show similar trends.

\subsection{Error Metrics}\label{sec:results-errors}

For field breakthrough curves, the true model parameters are unknown. Therefore, we assess the quality of the estimated parameters by comparing the breakthrough curve computed with those parameters (i.e., the reconstructed curve) with the observed data. To quantify the mismatch, we use two error metrics: the Root Mean Square Error (RMSE) and the Kullback–Leibler divergence (KLD). In both cases, we normalize the measured and simulated breakthrough curves so that their concentration values sum up to one, treating them as discrete probability distributions. The RMSE and KLD errors corresponding to any estimated set of parameters $\mathbf{y}$ are defined, respectively, as
\begin{linenomath*}\begin{align} \label{eq:RMSE}
    \varepsilon_{RMSE} (v,\mathbf{y})= \sqrt{\frac{1}{N^*}\sum_{i}^{N^*} \left(\frac{c(x^*,t_i,v,\mathbf{y}) \Delta^*_i}{\sum_{j}^{N^*}c(x^*,t_j,v,\mathbf{y})\Delta^*_j}-\frac{c^*_i\Delta^*_i}{\sum_{j}^{N^*}c^*_j\Delta^*_j}\right)^2},
    \\ 
    \label{eq:KL-div}
    \varepsilon_{KLD} (v,\mathbf{y}) =\sum_{i}^{N^*} \frac{c^*_i\Delta^*_i}{\sum_{j}^{N^*}c^*_j\Delta^*_j} \left( \ln\left(\frac{c^*_i\Delta^*_i}{\sum_{i}^{N^*}c^*_j\Delta^*_j}\right)+\ln\left(\frac{c(x^*,t_i,v,\mathbf{y})\Delta^*_i}{\sum_{i}^{N^*}c(x^*,t_i,v,\mathbf{y})\Delta^*_j}\right)\right),
\end{align}\end{linenomath*}
where $c(x,t,v,\mathbf{y})$ is the numerical solution at $(x,t)$ with parameters $v$ and $\mathbf{y}$. To ensure comparability of errors across different breakthrough curves, we use a common evaluation window  $t\in [0,24\cdot t_{\text{peak}}^*]$.

Because RMSE is based on squared differences, it aligns closely with the loss functions used in most of our optimization methods. For this reason, RMSE serves as our primary performance metric, while KLD is used as a secondary measure.

To evaluate if the synthetic dataset can be reused to estimate parameters of breakthrough curves that were not used to compute the distribution of the dataset, we split the breakthrough curves 90\% for training and 10\% for testing.

We compare the performance of the four methods within the DSTE framework (namely, DSTE-KL-NNI, DSTE-KL-PBI, DSTE-KL-FDNN, and DSTE-KL-IDNN) to Laplace domain fitting, moment matching and the global random optimization method (known as LIPO) with 1,000 forward evaluations per estimation. The four DSTE methods use a synthetic dataset of size 1,000. We also compare to the DSTE-KL-PBI results refined using LIPO with 300 evaluations per estimation and a narrower search space. The optimization constraints for this refinement are that the estimated velocity should be within 2\% of the DSTE-KL-PBI estimate and that the rest of the parameters should be within 30\% of the DSTE-KL-PBI estimate.

\begin{figure}[htb!]
	\centering
	\includegraphics[angle=0,width=\textwidth]{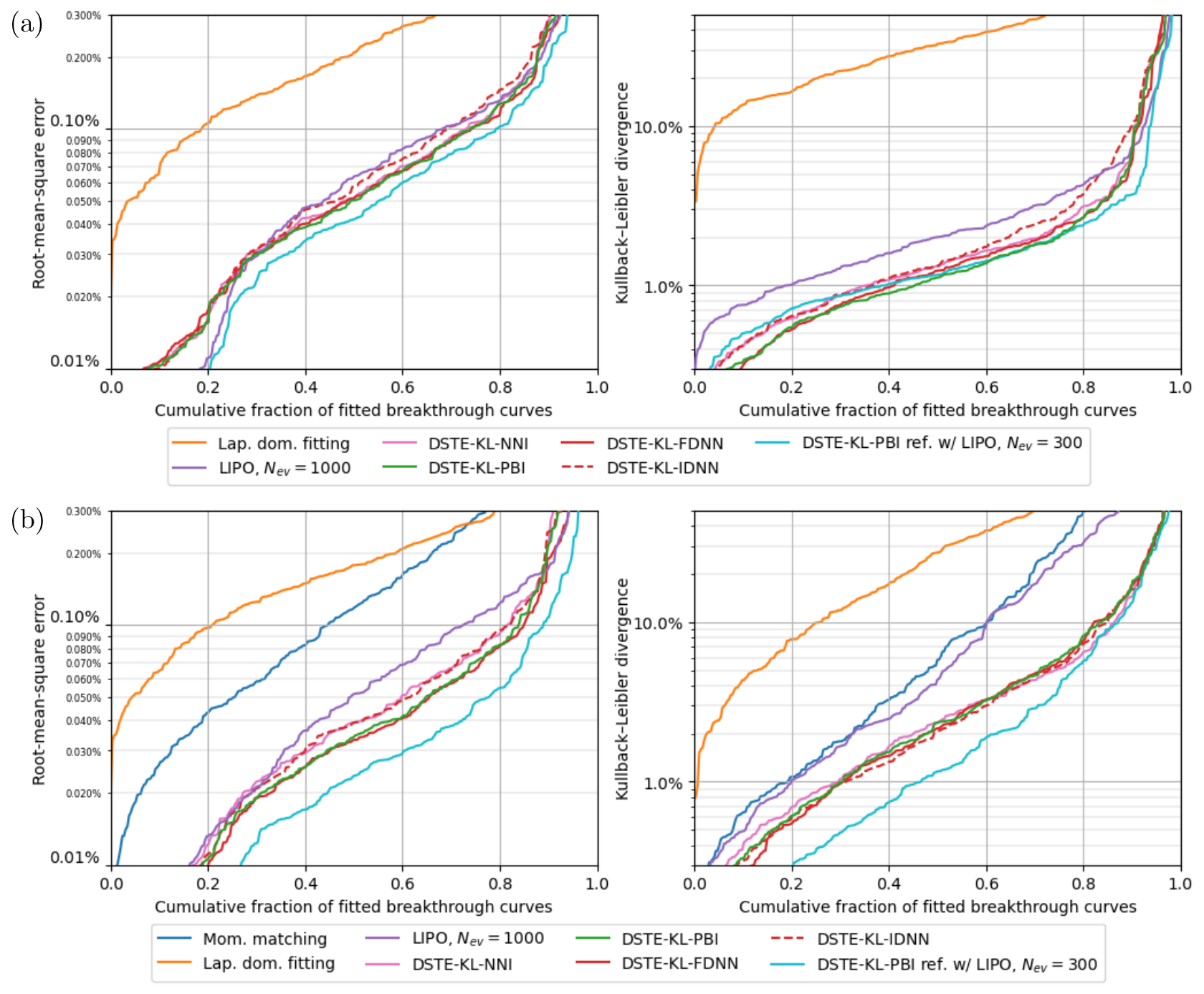}
	\caption{Comparison of RMSEs and Kullback-Leibler divergences of numerically reconstructed breakthrough curves (obtained with Laplace solver) using parameters estimated from different methods compared in this work. Errors are shown for all 295 measured breakthrough curves and are sorted from smallest to largest independently.  Domain, boundary and initial conditions from column 4 in Table \ref{tab:problems-comparison}. (a) First-order exchange model. (b) Power-law memory function model.}
	\label{fig:errors}
\end{figure}

Table \ref{tab:results} compares the accuracy of the seven methods and the refinement in estimating parameters in the advection-dispersion model with immobile exchange, described by first-order and power-law memory functions, for all 295 breakthrough curves. We summarize the results of the parameter estimation study using the median (since errors associated with different breakthrough curves span more than an order of magnitude, as seen in Figure \ref{fig:errors}) RMSE and KLD over the training or testing breakthrough curves. Additionally, the estimated parameters by each method are compared to the best estimate of the parameters, obtained by DSTE-KL-PBI refined with LIPO. The median of the absolute difference over all 295 breakthrough curves is shown.

\begin{table}
    \caption{Summary of results of parameter estimation from field breakthrough curves. Parameter estimates corresponding to the first-order exchange model and the power-law model. RMSEs and Kullback–Leibler divergence (KLD) are calculated for the breakthrough curves reconstructed with the Laplace forward solver (Sec. \ref{sec:methods-laplace-forward}) using the estimated parameters. Domain, boundary, and initial conditions from column 4 in Table \ref{tab:problems-comparison}. Errors from DSTE methods are averaged over 5 random realizations. They show a coefficient of variation around 15\%, due to the relatively small size of the synthetic dataset. Differences between them are consistent across different random realizations.}
    \centering
    \begin{tabular}{|>{\raggedright\arraybackslash}p{0.15\linewidth}|>{\centering\arraybackslash}p{0.08\linewidth}|>{\centering\arraybackslash}p{0.07\linewidth}|>{\centering\arraybackslash}p{0.07\linewidth}|>{\centering\arraybackslash}p{0.07\linewidth}|>{\centering\arraybackslash}p{0.08\linewidth}|>{\centering\arraybackslash}p{0.07\linewidth}|>{\centering\arraybackslash}p{0.07\linewidth}|>{\centering\arraybackslash}p{0.08\linewidth}|} \hline & Moms.& Lapl.&LIPO& DSTE-KL-NNI& DSTE-KL-PBI& DSTE-KL-FDNN&DSTE-KL-IDNN& DSTE-KL-PBI ref. w/ LIPO\\\hline\hline
 \multicolumn{9}{|c|}{\textbf{First-order exchange model}}\\\hline
 Error& \multicolumn{8}{|c|}{Median of errors}\\\hline
 RMSE (train)& 0.113\%& 0.169\%& 0.046\%&  0.037\%& 0.032\%& 0.032\%& 0.037\%& 0.022\%\\
 RMSE (test)& 0.167\%& 0.216\%& 0.102\%& 0.062\%& 0.053\%& 0.055\%& 0.057\% &0.038\%\\ \hline 
 KLD (train)& 5.27\%& 30.56\%& 4.40\%& 2.37\%& 2.38\%& 2.29\%& 2.17\%& 1.21\%\\
 KLD (test)& 8.72\% & 13.80\%& 5.32\%& 2.11\%& 1.84\%& 1.34\%& 1.19\%&0.64\%\\\hline
 Parameter& \multicolumn{8}{|c|}{Median differences with best estimated parameters (train \& test)}\\\hline
          $\left|\frac{\Delta v}{v}\right|$&  4.43\%& 2.69\%&
2.35\%& 1.14\%&  0.79\%& 
0.64\%&  0.79\%& \\ 
          $\left|\Delta \left(Pe^{-1}\right)  \right|$& 8.55e-4& 4.50e-4& 
5.20e-4&2.17e-4&  1.74e-4& 
1.64e-4&  1.81e-4& \\ 
          $\left| \Delta 
\left( \beta \hat{k}_f\right)  \right|$& 1.09& 0.708&
0.614& 0.292&  0.231& 
0.185&  0.198& \\ 
          $\left|\Delta \hat{k}_r  \right|$& 4.98& 23.8& 5.47&2.12&  1.75& 1.45&   1.54& \\\hline\hline
 \multicolumn{9}{|c|}{\textbf{Power-law model}}\\\hline
 Error& \multicolumn{8}{|c|}{Median errors}\\\hline
 RMSE (train)& & 0.200\%& 0.057\%& 0.050\%&  0.047\%& 0.049\%& 0.053\%& 0.040\%\\
 RMSE (test)& & 0.280\%& 0.102\%& 0.092\%& 0.090\%& 0.098\%& 0.106\%&0.084\%\\ \hline 
 
KLD (train)& & 33.07\%& 2.02\%& 1.36\%& 1.10\%& 1.24\%& 1.29\%& 1.17\%\\
 KLD (test)& & 23.02\%& 1.80\%& 1.07\%& 1.16\%& 1.22\%& 1.72\%&1.08\%\\\hline
 Parameter& \multicolumn{8}{|c|}{Median differences with best estimated parameters (train \& test)}\\\hline
 $\left|\frac{\Delta v}{v}\right|$& & 3.73\%& 1.41\%& 2.24\%& 1.83\%& 1.28\%& 1.83\%& \\ 
 $\left|\Delta \left(Pe^{-1}\right)  \right|$& & 8.23e-4& 3.77e-4& 2.60e-4& 1.94e-4& 1.89e-4& 2.52e-4& \\ 
 $\left|\Delta \left(\beta\hat{\alpha} \right)  \right|$& & 1.24e-1& 3.58e-2& 3.04e-2& 2.90e-2& 2.29e-2& 2.74e-2& \\ 
 $\left|\Delta \gamma   \right|$& & 5.74e-1& 4.86e-2& 8.20e-2& 5.13e-2& 4.87e-2& 4.47e-2& \\ \hline
    \end{tabular}
    \label{tab:results}
\end{table}

\begin{figure}[htb!]
	\centering
	\includegraphics[angle=0,width=\textwidth]{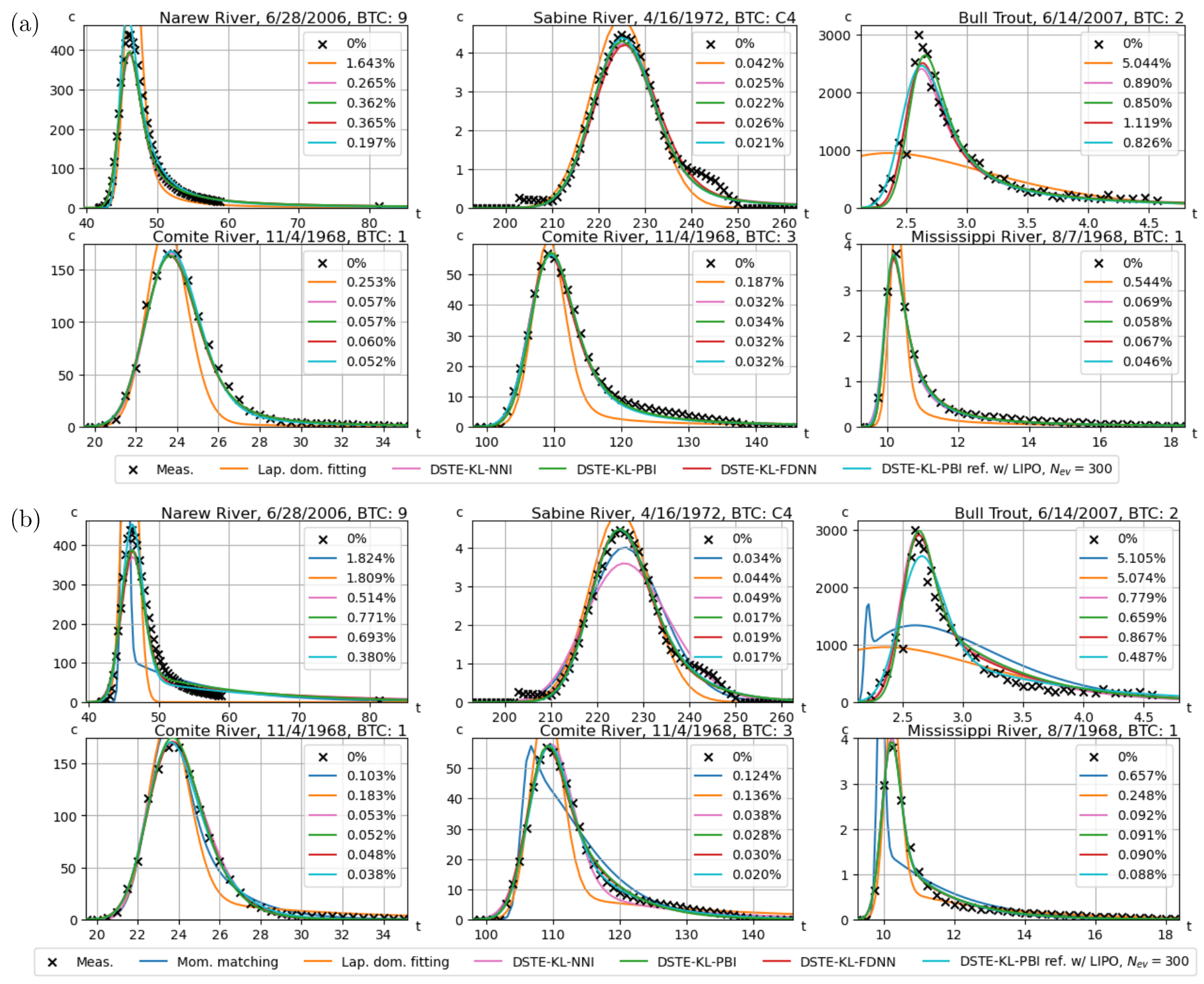}
	\caption{Comparison of measured breakthrough curves to reconstructed breakthrough curves from the advection-dispersion model with immobile exchange (obtained with forward Laplace solver) using parameters estimated with different methods. Domain, boundary and initial conditions from column 4 in Table \ref{tab:problems-comparison}. (a) First-order exchange model. (b) Power-law memory function model. The title of each breakthrough curve shows the river, the date of the tracer test, and the number of the breakthrough curve in the tracer test (a letter before the number signifies there were multiple tracer tests on the same day).}
	\label{fig:btcs}
\end{figure}

Figure \ref{fig:btcs} compares six randomly selected field breakthrough curves to those created numerically with parameters estimated from the five of the seven methods  (omitting LIPO and DSTE-KL-IDNN) and the refinement. Figure \ref{fig:errors} shows the RMSE and KLD errors in the seven methods and the refinement as functions of the cumulative fraction of the field breakthrough curves.

\subsection{Methods Not Requiring Forward Evaluations}

As shown in Figures \ref{fig:btcs} and \ref{fig:errors}, both Laplace domain fitting and moment matching yield mixed results in reconstructing breakthrough curves from their respective parameter estimates. These methods are highly sensitive to model misspecification, measurement noise, undersampling, and nonzero background concentrations. Such factors bias the least-squares fitting in transformed spaces (i.e., the Laplace or moment domain), which are not isometric to the original measurement space. Nevertheless, since these methods often produce reasonable approximations of the breakthrough curve, we use them to generate priors for constructing a synthetic dataset. However, we discard clear outliers—such as the Bull Trout sample from 6/14/2007 (2) shown in Figure \ref{fig:btcs}—following the procedure outlined in Section \ref{sec:methods-synthetic}.

Previous work by \citeA{seo_moment-based_2001} showed that Laplace domain fitting performs poorly under significant model misspecification, which aligns with our findings. However, our results suggest that the method can be effective when model mismatch is less severe.

\citeA{seo_moment-based_2001} and \citeA{gonzalez-pinzon_scaling_2013} have demonstrated the utility of moment matching for parameter estimation. However, direct comparisons with our approach are difficult due to differences in datasets (with only partial overlap), ambiguities in how they define and report error metrics and the limited number and precision of those metrics. Furthermore, our moment matching method is not directly comparable to either \citeA{seo_moment-based_2001} or \citeA{gonzalez-pinzon_scaling_2013}, since they supplement moment matching with additional constraints, either by incorporating prior knowledge or by combining it with least-squares fitting using random forward solutions, respectively. In contrast, we avoid using prior information in the Laplace domain fitting and moment matching methods to prevent potential bias. We also refrain from performing random forward evaluations, as these are more efficiently utilized within the DSTE framework. In our tests, the moment-matching objective functions proposed in earlier work underperformed relative to our own objective function (Eq \eqref{eq:moments-loss}) when used without these enhancements. Based on the comparison with the reconstructed breakthrough curves presented in their work, we conclude that DSTE-KL-PBI achieves better estimates than the moment-matching techniques.

\subsection{Comparison of DSTE Methods}

All four methods in the DSTE framework produced relatively accurate reconstructions of the breakthrough curves. NNI, the simplest of the methods, performed worse than PBI, although RMSEs and Kullback Leibler divergences were only 5 to 15\% larger. However, discrepancies in the estimated parameters were more pronounced, with errors 30–70\% larger than those from PBI or FDNN.

PBI and FDNN performed similarly, with a slight advantage to FDNN in training errors. PBI appears to show lower errors in the testing portion of the dataset. This may happen because the method will not extrapolate if parameters are out of range, rather it will project onto the synthetic dataset. A key drawback of the FDNN method is its computational cost, particularly during inversion defined by Eq \eqref{eq:KL-NN-inversion}. Unlike network training, inversion must be performed anew for each breakthrough curve, making it a performance bottleneck. Although FDNN provides better estimates of the advection velocity than PBI (because it jointly optimizes it with the rest of the parameters), it needs a good initial estimate for it, because, as mentioned in Section \ref{sec:methods-PBI}, the loss function has many local minima in the velocity component (this is a consequence of the dimensionless synthetic dataset and is not specific to any of the methods). 

IDNN with augmentation performed similar to or worse than NNI in terms of parameters. However, it sometimes showed the best results in terms of the Kullback Leibler divergence or in the estimation of some parameters. This highlights the fact that even though training an inverse neural network is an efficient way to estimate parameters, it is not clear in which sense the obtained parameters are optimal, since it does not minimize square residuals of the solution. Augmentation was fundamental for the performance of the method with the median RMSE for the first-order model being 0.052\% without augmentation, compared to 0.037\% with it, and for the power-law model being 0.070\% without augmentation, compared to 0.053\% with it.

\subsubsection{Application of Ready-Made Manifold Learning and Interpolation Techniques}

Additionally, we evaluated state-of-the-art manifold learning techniques, since manifold learning and interpolation have been successfully combined for parameter estimation in various contexts \cite{liu_integrated_2021, davidzon_cosmos2020_2022}. We use the implementations in \textit{scikit-learn} \cite{Pedregosa_Scikit-learn_Machine_Learning_2011} of ISOMAP \cite{tenenbaum_global_2000} and Locally Linear Embedding (LLE) \cite{roweis_nonlinear_2000}, to generate $d$-dimensional embeddings of the synthetic data manifold. These embeddings were intended to enable interpolation in a space matching the intrinsic dimensionality of the manifold. We then applied linear interpolation using the \textit{LinearNDInterpolator} from \textit{scipy.interpolate} to estimate parameters in this embedding space. However, both ISOMAP and LLE performed worse than NNI, and we identify two main reasons for this.

The first issue occurs in the step where the methods embed the manifold onto the $d$-dimensional space, known as Multidimensional Scaling. The transformation deforms the manifold because it cannot be isometric (a general manifold requires $d+1$ dimensions for an isometric immersion \cite{kuiper_c1-isometric_1955}, i.e., an embedding that allows self crossings, but manifold learning methods usually require even more dimensions). Multidimensional scaling could be done onto a higher-dimensional space, in which isometry may be better preserved, but we would still need to project measurements onto the synthetic manifold to avoid extrapolation, defeating the purpose of manifold learning.

The second issue with these manifold learning methods occurs when they transform measured points onto the $d$-dimensional space. If the points are far from the manifold, the projection is poor. We observe this by the fact that, for our measured dataset, on average, around 2.5 out of four nearest neighbors in KL space remain as part of the four nearest neighbors in the transformed space for both methods. This hurts the quality of interpolation, since the four nearest neighbors are a close proxy for interpolating nodes in linear interpolation (though they are not the same as \textit{LinearNDInterpolator} uses Delaunay triangulation).

\subsection{Comparison to Global Optimization Method and Refinement of PBI Results}\label{sec:results-LIPO}

Table \ref{tab:results} shows that LIPO consistently performs worse than DSTE-KL-PBI, though the magnitude of the performance gap varies. This variation appears to be linked to the characteristics of the prior distribution of parameters. In first-order exchange—where the parameter distribution closely resembles a multivariate lognormal—DSTE-KL-PBI outperforms LIPO by a larger margin than in the power-law model, where the lognormal approximation is poorer. KLDs show a larger difference in favor of DSTE methods, suggesting LIPO is more likely to overfit to its loss function.

The performance difference is particularly notable when considering computational costs. LIPO requires 1,000 forward simulations per parameter estimation (i.e., per breakthrough curve), whereas the synthetic dataset is generated only once and reused across all 295 estimations. Moreover, the same synthetic dataset generalizes to data not used in the estimation of the prior distribution, as demonstrated by good testing errors in DSTE-KL-PBI.

In our specific case, where the forward model is computationally inexpensive, we can afford the 1,000 forward evaluations per curve required by LIPO. However, for problems involving even moderately more expensive forward models, LIPO would quickly become computationally infeasible. In contrast, our synthetic dataset methods can be applied to any transport equation that can be expressed in a dimensionless form, making them more scalable and suitable for such problems.

That said, LIPO remains valuable as a refinement tool. Both its global stochastic search and local optimization steps can effectively refine parameter estimates if a good initial guess is available and the search space is sufficiently constrained. For instance, in the refinement setup we use, the effective size of the search space $(V(\chi))^{1/(d+1)}$ (which defines the scaling of the loss) is approximately 20 times smaller than in the original LIPO optimization for the first-order model and 3 times smaller for the power-law model.

\subsection{Other Comparisons}\label{sec:results-other_comparison}

For completeness, we also compared our methodology to local optimization methods. We used optimization algorithms implemented in\textit{ scipy optimize} \cite{2020SciPy-NMeth}, such as Nelder-Mead, Powell, and L-BFGS-B. None of them improved significantly on an initial guess obtained by DSTE-KL-PBI, and often, they didn't converge.

\subsection{Special Considerations for the Power-Law Model}

From the Laplace transform of the power-law memory function (Table \ref{tab:memory_functions}) and the Laplace domain solution (Eq \eqref{eq:laplace-solution}), we see that as $\gamma\to1$, the solution is the same as the pure advection-dispersion equation with an effective velocity of $\frac{v}{1+\beta\hat{\alpha}}$. For this reason, when the value of $\gamma$ is close to $1$, it often happens that the value of $v$ that minimizes RMSEs is much higher than the velocity of the peak $\frac{L}{t^*_{\text{peak}}}$, to the level of being unphysical.

We consider such estimates unphysical based on two observations: (1) the estimated velocities for the first-order model and the pure advection-dispersion equation are typically within 20\% of each other, and (2) where mean channel velocities are known, they tend to lie within a factor of two of the estimated advection velocities. In contrast, in the power-law model, the estimated velocities can be more than four times higher than these. 

For this reason, in the power-law model, we regularize the loss functions with a term $\lambda_{reg}\left(v-\frac{L}{t^*_{\text{peak}}}\right)^2$. The value of $\lambda_{reg}$ depends on the characteristics of the rest of the loss function, but it is chosen to be such that $v$ deviates from the velocity of the peak by 4\% on average, to match the mean obtained in the first-order model.

\subsection{Application of the Method to New Tracer Tests}

Training and testing errors in Table \ref{tab:results} show that DSTE-KL-PBI is applicable to breakthrough curves not used to estimate the distribution of parameters. While testing errors are higher than training errors, DSTE-KL-PBI exhibits the lowest training-to-testing error ratios among all methods. This reflects its robustness across diverse breakthrough curves. The higher testing errors are due to some breakthrough curves being harder to fit than others (explained by the fact that errors span several orders of magnitude, as seen in Figure \ref{fig:errors}).

\subsection{Scaling Behavior and Practical Trade-offs of Parameter Estimation Methods}

In the DSTE framework, the precision of the interpolation step is determined by the length $h$ of the edges in the simplices forming the KL-space manifold, which scales as  $h=\mathcal{O}\left(N_{\text{synth}}^{-1/d}\right)$. Assuming the loss is defined as an L2-norm error metric (e.g., RMSE) and excluding irreducible error from model misspecification, we can derive theoretical scaling laws for the loss as a function of the training set size. 

For NNI, the loss scales as $\mathcal{O}\left(h\right)=\mathcal{O}\left(N_{\text{synth}}^{-1/d}\right)$, since the error is the distance to the nearest point, which is proportional to $h$. In contrast, PBI uses local linear interpolations in KL space (each barycentric coordinate is a linear function of the KL space coordinates), leading to a second-order scaling $\mathcal{O}\left(h^{2}\right)=\mathcal{O}\left(N_{\text{synth}}^{-2/d}\right)$\cite{wendland_local_2004}. 

The loss in Neural Networks scales with the amount of data and the dimensionality of the input according to a power-law \cite{henighan_scaling_2020, hestness_deep_2017, hoffmann_training_2022}. If DSTE-KL methods are consistent with these trends, the scaling should be $\mathcal{O}\left(N_{\text{synth}}^{-\alpha}+N^{-\beta}+L_0\right)$ for FDNN and $\mathcal{O}\left(\left(N_{aug}N_{\text{synth}}\right)^{-\alpha}+d^{-\beta}+L_0\right)$ for IDNN, where $L_0$ is the irreducible loss due to limited model representation capacity, and empirical studies suggest  $\alpha\approx\beta\approx 0.3$. Other interpolation schemes, such as Gaussian Radial Basis Function interpolation, offer better asymptotic behavior with exponential convergence $\mathcal{O}\left(\exp\left(-\frac{c |\log(h)|}{h}\right)\right)=\mathcal{O}\left(\exp\left(-\frac{c\log(N_{\text{synth}})}{dN_{\text{synth}}^{-1/d}}\right)\right)$, but were not pursued due to the need for hyperparameter tuning of the shape parameter (which should be variable for non-uniform spacing of the points of the training dataset). We remind that LIPO's losses scale like $\mathcal{O}\left(N_{\text{ev}}^{-1/(d+1)}\right)$.

For our case with \( d = 3 \), we observe the following ranking in scaling efficiency: LIPO (\( \alpha = 0.25 \)) $<$ DSTE-KL-DNN (\( \alpha \approx 0.3 \)) $<$ DSTE-KL-NNI (\( \alpha = 0.33 \)) $<$ DSTE-KL-PBI (\( \alpha = 0.66 \)). DSTE-KL-DNN only starts to scale better than PBI for $d\geq7$. This may be too many parameters to identify in a single breakthrough curve, though combining information from multiple breakthrough curves and other sources can make this possible.

It is important to note that these scaling laws describe asymptotic behavior as \( N_{\text{synth}} \to \infty \). In real applications where forward evaluations are costly and dataset sizes are limited, lower-order terms and constant factors may dominate, and practical performance may diverge from theoretical expectations.

We compare the scaling and other characteristics for each method in Table \ref{tab:methods-characteristics}. This comparison is not exhaustive. Other relevant differences include differentiability of the loss functions, computational costs not associated with forward evaluations, generality of the methods, ease of implementation, etc.
\begin{table}
    \centering
     \caption{Comparison of characteristics of different parameter estimation methods.}
    \begin{tabular}{|>{\raggedright\arraybackslash}p{0.14\linewidth}|>{\raggedright\arraybackslash}p{0.2\linewidth}|>{\raggedright\arraybackslash}p{0.15\linewidth}|>{\raggedright\arraybackslash}p{0.17\linewidth}|p{0.18\textwidth}|} \hline 
         Method&  Requirements &Type of estimate&  Sources of error&Scaling of losses with the number of forward evaluations.\\\hline
         Fit statistics or  sol. in transformed domain.&  Expressions for statistics or solution in transformed domain.&Least-square on statistics or transformed domain.&  Errors in computing the expression and regression error.&No forward evaluations needed.\\\hline
         Global optimization&  Forward solver and bounds. &Least-square.&  Fwd. sol., regression.&$\mathcal{O}\left(N_{\text{ev}}^{-1/(d+1)}\right)$\\ \hline 
 DSTE-KL-NNI& Forward solver and training set.& Least-square.& Fwd. sol., interpolation.&$\mathcal{O}\left(N_{\text{ev}}^{-1/d}\right)$\\\hline
  DSTE-KL-PBI& Fwd. solver and training set.&Least-square.& Fwd. sol., projection, interpolation.&$\mathcal{O}\left(N_{\text{ev}}^{-2/d}\right)$\\\hline
         DSTE-KL-FDNN&  Fwd. solver, training set, and hyperparameter tuning.&Least-square.&  Fwd. sol., approximation, representation, regression.&$\mathcal{O}\left(N_{\text{ev}}^{-\alpha}\right.$ $\left.+N^{-\beta}+L_0\right)$\\\hline
         DSTE-KL-IDNN&  Fwd. solver, training set, estimate of projection vectors, and hyperparameter tuning.&Inverse least-square.&  Fwd. sol., approximation, representation, interpolation.&$\mathcal{O}\left(N_{\text{ev}}^{-\alpha}\right.$ $\left.+d^{-\beta}+L_0\right)$\\ \hline
    \end{tabular}

    \label{tab:methods-characteristics}
\end{table}

\section{Analysis of Estimated Parameters}\label{sec:data-analysis}

In this section, we analyze the parameters obtained by DSTE-KL-PBI refined using LIPO for the 295 breakthrough curves of the TIERRAS dataset.

\subsection{Comparison of Different Models}

\begin{figure}[htb!]
	\centering
	\includegraphics[angle=0,width=\textwidth]{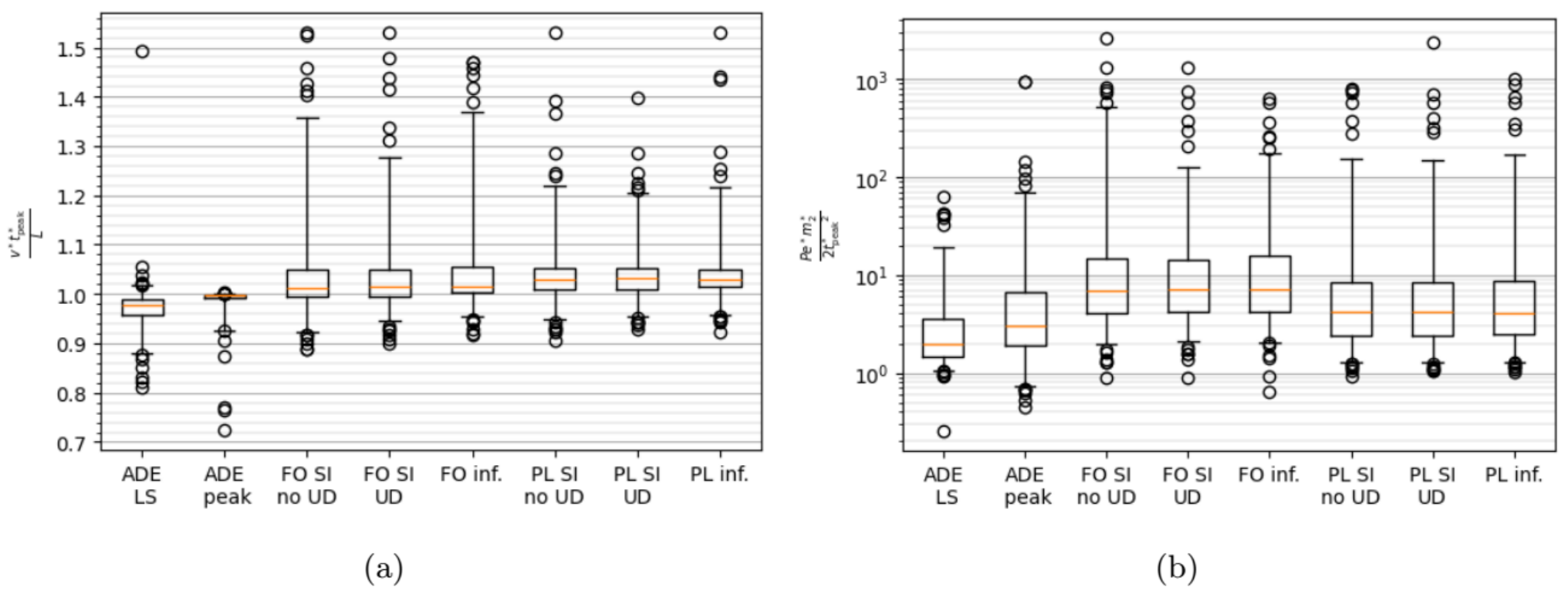}
	\caption{Comparison of velocity and Peclet number fitted for different models to 295 breakthrough curves from TIERRAS (percentiles 2, 25, 50, 75, and 98). (ADE peak) and (ADE LS) fit the advection-dispersion equation in an infinite domain to the peak or to the whole breakthrough curve by least-squares regression, respectively, Section \ref{sec:ADE-fit}. The rest of the parameters are obtained by DSTE-KL-PBI refined by LIPO for first-order exchange (FO) or power-law (PL) models, in an infinite domain (inf.), semi-infinite domain with upstream dispersion (SI UD), or semi-infinite domain without upstream dispersion (SI no UD).}
	\label{fig:params_comparison}
\end{figure}

Figure \ref{fig:params_comparison}, summarizes the estimated velocity and Peclet number (the two parameters common to all models) for different models and boundary conditions. We compare the ADE model with immobile exchange, described by both memory functions and three definitions of the boundary conditions. We also compare them with the parameters obtained by fitting the ADE (without immobile exchange) to the peak of the breakthrough curve and the whole breakthrough curve. The parameters are normalized to facilitate comparison. We normalize $v^*$ with $\frac{L}{t^*_{\text{peak}}}$ and $Pe^*$ with $\frac{2 {t^*_{\text{peak}}}^2}{m_2^*}$. These are used for normalization because, as $Pe\to\infty $, $\frac{v^*t^*_{\text{peak}}}{L}\to1$ and $\frac{Pe^* m_2^*}{2 {t^*_{\text{peak}}}^2}\to 1$ in the ADE solution.

Results show that the parameter estimates differ significantly depending on the inclusion of immobile exchange. On average, the Peclet number is about twice as large when using memory-function models compared to the ADE fitted at the peak, and approximately 2.5 times larger than the ADE fitted to the entire breakthrough curve. This suggests that, if memory-function models accurately represent physical processes, more than half of the apparent dispersion in river systems can be attributed to exchange with the immobile phase. Even when tails are ignored (as in peak fitting), immobile exchange still accounts for a substantial fraction of dispersion.

Phase exchange also delays tracer transport, which explains the larger velocities estimated with memory-function models. The interplay between advection velocity and transient storage parameters—and its effect on tracer travel time—has been previously discussed in \citeA{runkel_new_2002}. Comparing estimated advection velocities to measured channel velocities may provide a validation path for these models, though care must be taken due to variability within reaches.

In contrast, the choice of domain definition (columns 1–4 in Table \ref{tab:problems-comparison}) has a negligible effect on RMSEs, with differences under 2\%. This suggests that breakthrough curves alone do not favor one domain assumption over another. The close similarity in error is likely due to compensatory shifts in parameter estimates. For instance, in the first-order model, estimated velocities are on average 0.4\% higher under the semi-infinite domain with upstream dispersion than without it, and 1.1\% higher under the infinite domain. The inverse Peclet number (a dimensionless form of the dispersion coefficient) remains consistent between the semi-infinite domains but is, on average, slightly lower (by 
$3\times10^{-4}$) in the infinite domain. Analytical solutions of the ADE without immobile exchange (last row of Table \ref{tab:problems-comparison}) indicate that, for identical parameters, peak arrival times are delayed and variance reduced in an infinite domain relative to a semi-infinite one. Consequently, to match observed breakthrough curves, parameter estimates in the infinite domain reflect earlier peaks (via higher velocities) and wider spreads (via higher inverse Peclet numbers). These effects are less clear in the power-law model due to higher fitting errors and regularization of the velocity.

\subsection{Trends in the Parameters}

\begin{figure}[htb!]
	\centering
	\includegraphics[angle=0,width=\textwidth]{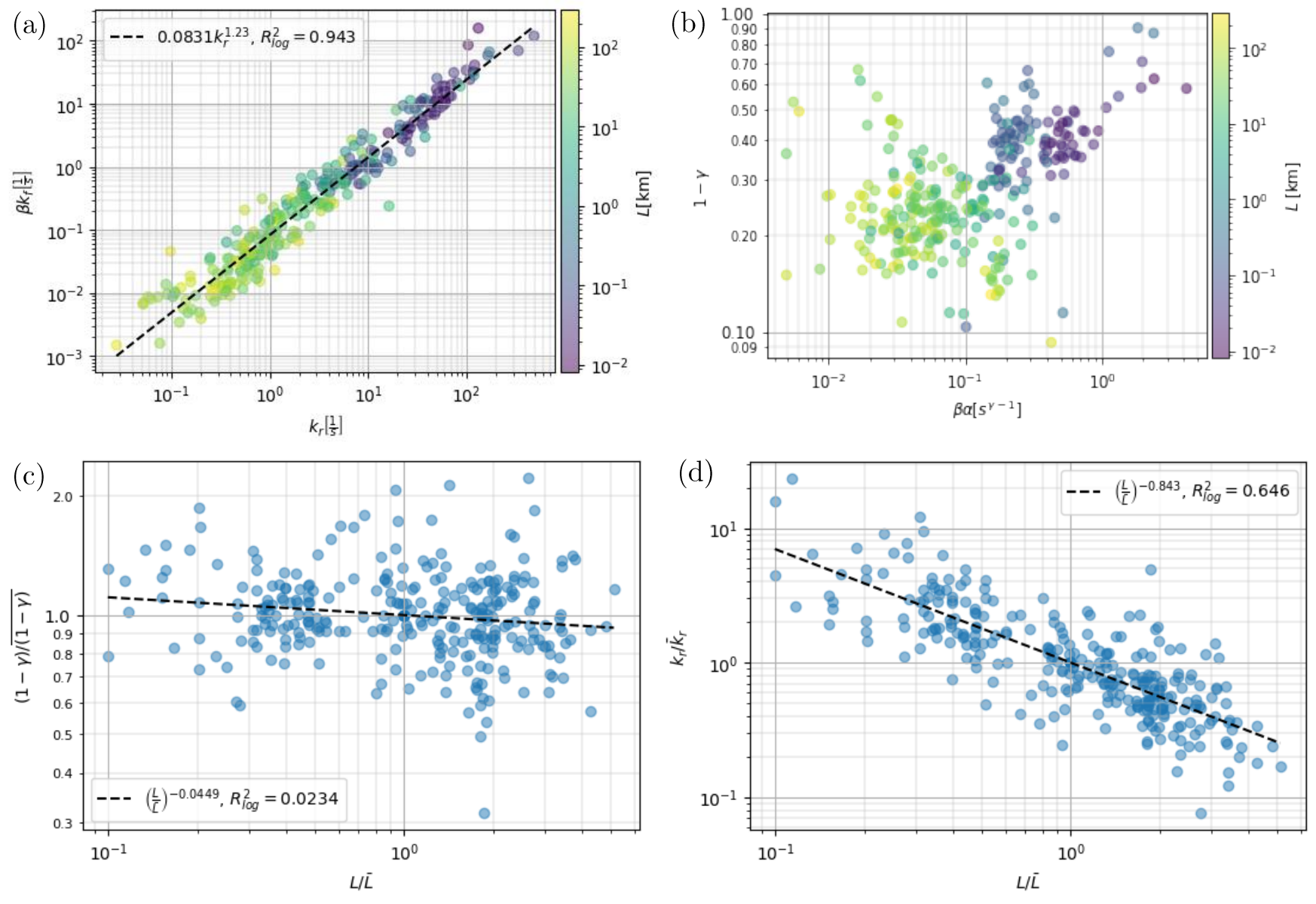}
	\caption{Summary of fitted parameters of the memory functions and relations between them for 295 breakthrough curves of the TIERRAS dataset. Dashed lines are power-law fits. (a) Inverse rate $k_r$ vs. forward rate $\beta k_f$, parameters of the first-order exchange model. The color scale represents space dependence. (b) Exchange scale $\beta \alpha$ vs power-law exponent $\gamma$, parameters of the power-law model. The color scale represents space dependence. (c) $\gamma$ versus $L$. (d) $k_r$ versus $L$. In (c) and (d), $\bar{L}$, $\bar{k}_r$, and $\overline{1-\gamma}$ represent the geometric means of $L$, $k_r$, and $1 - \gamma$, respectively, calculated across all breakthrough curves measured in the same tracer experiment (at different locations).}
	\label{fig:params}
\end{figure}

Figure \ref{fig:params} summarizes the fitted parameters of the memory functions (the parameters that govern the exchange between the mobile and immobile phases). Parameters for the first-order memory function are shown in Figures \ref{fig:params}a and d, and those for the power law are displayed in Figures \ref{fig:params}b and c. In these figures, each point represents one of the 295 field breakthrough curves from the TIERRAS dataset. Figures \ref{fig:params}a and b compare the pair of parameters that define either memory function. Figures \ref{fig:params}c and d show how, in each case, one of these parameters depends on the distance from the injection point. Figures \ref{fig:params}c and d normalize the distances and parameters with their corresponding geometric mean calculated across all breakthrough curves measured in the same tracer experiment (breakthrough curves measured at different downstream distances from the same release are normalized independently, but plotted together).

In Figure \ref{fig:params}a, we see that the fitted parameters are correlated to each other, with $\beta k_f$ being roughly proportional to $k_r$. This finding supports the assumption of the reversible exchange (i.e., that $\beta$ is constant and $k_f=k_r$) in the Transient Storage Model  (TSM). However, the assumption cannot be confirmed without direct measurement of some of the parameters or breakthrough curves in the immobile phase.

The color scale of Figure \ref{fig:params}a shows that the value of $k$  depends on $L$, the distance from the tracer release to the location of the breakthrough curve. Figure \ref{fig:params}d, shows a power-law trend between distances and exchange rates $k$. In this figure, the parameters and distances are normalized independently within each tracer test to highlight trends within the experiment. Without this normalization, the overall pattern is less apparent due to variations in discharge and physical settings across experiments. Similar strong dependence of the parameters on the length scale was observed in heterogeneous porous media transport models \cite{FERNANDEZGARCIA2005745}.

In the case of the power-law model, Figure \ref{fig:params}b shows that estimated values of $\beta\hat{\alpha}$ and $\gamma$ are not well correlated and that $\gamma$ approaches one as $L$ increases. However, this trend is only visible when comparing different tracer tests. Figure \ref{fig:params}c doesn't show a significant trend within tracer tests.

Space variations are consistent across experiments. In first-order exchange, the exchange rate scales as $k\sim L^{-0.84}$. The exchange rate coefficient scaling is consistent with the scaling of the temporal coefficient of skewness $\tilde{\mu}_3$. According to \cite{aghababaei_temporal_2023}, as $Pe\to\infty$, skewness is related to the exchange parameters by $(\tilde{\mu}_3^2L)^{-1}\propto\beta k$. We observe a similar scaling $(\tilde{\mu}_3^2L)^{-1}\sim L^{-0.74}$, in the measured temporal moments, which indicates our observations are not an artifact of the estimation method. Interestingly, previous studies \cite{Scott2003-yf} found that, for much shorter tracer tests (breakthrough curves measured at distances up to 619 m, compared to the 592 m to 293 km range in our dataset), the scaling is reversed (the exchange rate increases with distance). The behavior they observe is consistent with preasymptotic behavior of a release in the center of the stream. Preasymptotic behavior happens when the section is not fully mixed and effective transport parameters vary, converging to an asymptotic value as time goes to infinity. Under a velocity profile with higher velocities in the center and a central release, $(\tilde{\mu}_3^2L)^{-1}$ grows until it reaches an asymptotic value.

In contrast, only 16\% of our breakthrough curves are sampled before the asymptotic regime is reached, that is, before we get a fully mixed regime in the river section. For this to happen, according to \cite{fischer_chapter5_1979}, we need to reach a downstream distance of around $0.2 L_h$, where $L_h=\frac{UB^2}{d_H u_{*}}$, $U$ is the mean velocity of the channel, $B$ is the width, $d_H$ is the hydraulic depth, and $u_{*}$ is the shear velocity. Some of the scaling we observe may be attributed to increases in discharge in the downstream direction. However, only a few tracer tests show very significant differences. The scaling may then be caused by different processes that are not captured by the memory function models, such as a coupling between immobile phase exchange and preasymptotic dispersion (each exchange into the mobile zone could be considered as a release at the boundaries of the stream, i.e., the bed or the banks, with its corresponding variations of the transport parameters, expanding on the work by \citeA{Gurung2024-bl}) or lengthwise variations of velocity profiles. 

\section{Conclusions}\label{sec:conclusions}

This work advances river transport modeling by introducing a novel framework for data-model integration, parameter estimation, and model evaluation, addressing several gaps in the state-of-the-art methods. Specifically,  
we developed the DSTE framework for the decoupled estimation of the velocity and dimensionless parameters, which greatly reduces the need for forward evaluations in inverse modeling of transport processes. We demonstrated that the DSTE framework is more accurate than the state-of-the-art methods for estimating parameters in several river transport PDE models.

We generated a reusable low-dimensional synthetic dataset applicable to new observations beyond the training field data. 

We formulated a novel Projected Barycentric Interpolation method in the KL space for estimating parameters using the synthetic dataset.

We extended Laplace-domain solutions to accommodate different domain definitions and memory functions.

We generalized analytical solutions for temporal moments to different boundary conditions.

We improved the augmentation technique for inverse Neural Networks learning of a compact manifold.

We performed a comparative analysis of different parameter estimation methods.

We estimated parameters for the tracer experiments in the TIERRAS dataset, which is a significant enhancement of this dataset. For its 295 breakthrough curves, we provided parameters for several transport models. In addition to breakthrough curves, this dataset includes characteristics of the tracer experiments and hydraulic parameters of the rivers and streams. Therefore, the enhanced dataset can be used to relate the transport parameters (estimated in this paper) to hydrological parameters and the characteristics of the tracer experiments using data analysis and machine learning methods. 

Future research should focus on uncertainty quantification and a deeper understanding of the dependence of parameters on the model's length scale. We hypothesize that the latter can be achieved by linking smaller-scale processes to river reach transport and the physical properties of the rivers. This hypothesis can be tested by incorporating more complexity into one-dimensional models and comparing their results with those from two and three-dimensional models.

\section{Open Research}

All data used for this research are available at \url{https://www.tierras.org/}. The codes used in this study are openly available in a GitHub repository at \url{https://github.com/manuelmreyna/inverse-river-transport}. The repository includes:
\begin{itemize}
    \item A Python implementation of the DSTE-KL-PBI method.
    \item Pre-trained synthetic datasets that can be used to estimate parameters using DSTE-KL-PBI. These training datasets were created with $\ln\left(\mathbf{y}\right)\sim \mathcal{N}_d\left(\boldsymbol{\mu}_{log,prior},b\Sigma_{log,prior}\right)$, with $b=4$, ensuring that they cover a wider range of parameters than that observed in the TIERRAS dataset. As a result, the prebuilt datasets are suitable for fitting tracer tests that differ substantially from those used in the original parameterization. The key advantage of this setup is that it allows parameter estimation without requiring any additional forward simulations.
    \item Sets of parameter estimates for the 295 breakthrough curves from the TIERRAS dataset analyzed in this work. These estimates were obtained by DSTE-KL-PBI and refined by LIPO for the transport models with the two memory functions.
    \item Two Jupyter notebooks that reproduce the results and figures presented in this paper, including the parameter estimation workflow and performance evaluation.
\end{itemize}

\section{Acknowledgements}
This research was supported by the United States National Science Foundation and the U.S. Department of Energy (DOE) Advanced Scientific
Computing Research program. Pacific Northwest National Laboratory is
operated by Battelle for the DOE under Contract DE-AC05-76RL01830.

\bibliography{references}

\end{document}